\documentclass[aps,prd,12pt,notitlepage,showpacs,nofootinbib,tightenlines]{revtex4-1}
\usepackage{amsmath}
\usepackage{amssymb}
\usepackage{epsfig}
\usepackage{graphicx}
\usepackage{bm}
\usepackage{times}
\usepackage{braket}
\usepackage{color}
\usepackage{slashed}
\usepackage{hyperref}
\definecolor{Red}{rgb}{1.,0.,0.}

\definecolor{Blue}{rgb}{0.,0.,1.}

\definecolor{nicered}{rgb}{0.7,0.1,0.1}
\definecolor{nicegreen}{rgb}{0.1,0.5,0.1}
\hypersetup{colorlinks,citecolor=nicegreen,linkcolor=nicered}

\begin{document}
\newcommand{\beq}{\begin{eqnarray}}
\newcommand{\eeq}{  \end{eqnarray}}
\newcommand{\ben}{\begin{enumerate}}
\newcommand{\een}{  \end{enumerate}}
\newcommand{\non}{\nonumber\\ }

\newcommand{\jpsi}{J/\Psi}
\newcommand{\ppa}{\phi_\pi^{\rm A}}
\newcommand{\ppp}{\phi_\pi^{\rm P}}
\newcommand{\ppt}{\phi_\pi^{\rm T}}
\newcommand{\ov}{ \overline }
\newcommand{\zerot}{ {\textbf 0_{\rm T}} }
\newcommand{\kt}{k_{\rm T} }
\newcommand{\kta}{{\textbf k_{\rm 1T}} }
\newcommand{\ktb}{{\textbf k_{\rm 2T}} }
\newcommand{\ktc}{{\textbf k_{\rm 3T}} }
\newcommand{\fb}{f_{\rm B} }
\newcommand{\fk}{f_{\rm K} }
\newcommand{\rk}{r_{\rm K} }
\newcommand{\mb}{m_{\rm B} }
\newcommand{\mw}{m_{\rm W} }
\newcommand{\im}{{\rm Im} }
\newcommand{\kks}{K^{(*)}}
\newcommand{\acp}{{\cal A}_{\rm CP}}
\newcommand{\pb}{\phi_{\rm B}}
\newcommand{\xeba}{\bar{x}_2}
\newcommand{\xsba}{\bar{x}_3}
\newcommand{\peas}{\phi^A}
\newcommand{\pvsl}{ p \hspace{-2.0truemm}/_{K^*} }
\newcommand{\esl}{ \epsilon \hspace{-2.1truemm}/ }
\newcommand{\psl}{ p \hspace{-2truemm}/ }
\newcommand{\ksl}{ k \hspace{-2.2truemm}/ }
\newcommand{\lsl}{ l \hspace{-2.2truemm}/ }
\newcommand{\nsl}{ n \hspace{-2.2truemm}/ }
\newcommand{\vsl}{ v \hspace{-2.2truemm}/ }
\newcommand{\epsl}{\epsilon \hspace{-1.8truemm}/\,  }
\newcommand{\bfkk}{{\bf k} }
\newcommand{\calm}{ {\cal M} }
\newcommand{\calh}{ {\cal H} }
\newcommand{\calo}{ {\cal O} }

\def \appb{{\it Acta. Phys. Polon. \bf B }  }
\def \cpc{ {\it Chin. Phys. \bf C } }
\def \csb{ {\it Chin. Sci. Bull.} }
\def \ctp{ {\it Commun. Theor. Phys. } }
\def \epjc{{\it Eur. Phys. J. \bf C} }
\def \jhep{{\it JHEP } }
\def \jpg{ {\it J. Phys. G} }
\def \mpla{{\it Mod. Phys. Lett. \bf A } }
\def \npb{ {\it Nucl. Phys. \bf B} }
\def \plb{ {\it Phys. Lett. \bf B} }
\def \ppnp{{\it Prog.Part. Nucl. Phys. } }
\def \pr{  {\it Phys. Rep.} }
\def \prc{ {\it Phys. Rev. \bf C }}
\def \prd{ {\it Phys. Rev. \bf D} }
\def \prl{ {\it Phys. Rev. Lett.}  }
\def \ptp{ {\it Prog. Theor. Phys. }}
\def \zpc{ {\it Z. Phys. \bf C}  }
\def \ap{ {\it Ann. of Phys}  }
\def \rmp{ {\it Rev.Mod.Phys. }  }

\title{\boldmath The NLO contributions to the scalar pion form factors
and the ${\cal O}(\alpha_s^2)$ annihilation corrections to the $B\to \pi\pi$ decays}
\author{Shan Cheng$^{1}$} \email{chengshan-anhui@163.com}
\author{Ya-Lan Zhang$^{1}$} \email{644534413@qq.com}
\author{Zhen-Jun Xiao$^{1,2}$ } \email{xiaozhenjun@njnu.edu.cn}
\affiliation{1.  Department of Physics and Institute of Theoretical Physics,
Nanjing Normal University, Nanjing, Jiangsu 210023, People's Republic of China,}
\affiliation{2. Jiangsu Key Laboratory for Numerical Simulation of Large Scale Complex Systems,
Nanjing Normal University, Nanjing 210023, People's Republic of China}
\date{\today}
\vspace{1cm}
\begin{abstract}
In this paper, by employing the $k_{T}$ factorization theorem,
we made the first calculation for the space-like scalar pion form factor
$Q^2 F(Q^2)$ at the leading order (LO)
and the next-to-leading  order (NLO) level, and then found the time-like
scalar pion form factor $F'^{(1)}_{\rm a,I}$
by analytic continuation from the space-like one.
From the analytical evaluations and the numerical results, we found the following points:
(a) the NLO correction to the space-like scalar pion form factor  has an opposite sign with the LO
one but is very small in magnitude, can produce at most $10\%$ decrease to LO result
in the considered $Q^2$ region;
(b) the NLO time-like scalar pion form factor  $F'^{(1)}_{\rm a,I}$
describes the ${\cal O}(\alpha_s^2)$ contribution to the factorizable annihilation diagrams
of the considered $B \to \pi\pi$ decays, i.e. the NLO annihilation correction;
(c) the NLO part of the form factor $F'^{(1)}_{\rm a,I}$ is very small in size, and
is almost independent with the variation of cutoff scale $\mu_0$, but this form factor
has a large strong phase around $-55^\circ$ and may play an important role in producing large
CP violation for $B\to \pi\pi$ decays;
and (d) for $B^0 \to \pi^+\pi^-$ and $ \pi^0\pi^0$ decays, the newly known
NLO annihilation correction can produce only a very small enhancement
to their branching ratios, less than $3\%$ in magnitude, and therefore we could not interpret
the well-known $\pi\pi$-puzzle by the inclusion of this NLO correction to the factorizable annihilation
diagrams.
\end{abstract}



\maketitle

\section{Introduction}
As an important application of the $\kt$ factorization
theorem \cite{plb242-97,npb325-62,npb360-3,npb366-135,npb381-129,zpc50-139,prd52-5358},
the perturbative QCD (pQCD)  factorization approach
\cite{prd53-2480,prd63-074009,prd65-014007,epjc23-275,prd66-094010}
has been widely used to deal with various $B/B_s$ meson decays for example in
Refs.\cite{prd72-114005,prd86-114025,prd87-094003,prd90-014029,csb59-125}.
One advantage of the pQCD factorization approach is that the annihilation diagrams in the heavy-to-light
decays are calculable\cite{prd67-034001}. In pQCD approach, the annihilation
diagrams can provide a large strong phase which is essential to generate the large
CP violation for some $B/B_s$ meson decay channels.

In recent years, the $\kt$ factorization theorem is greatly improved after intensive studies
by many authors. By using the universal gauge invariant wave functions with the inclusion of
high order contributions \cite{prd64-014019,epjc40-395,jhep0601-067},
the next-to-leading order (NLO) corrections to the form factors for some transition processes
have been calculated \cite{prd76-034008,prd83-054029,prd85-074004,prd89-054015,prd89-094004}
during the past decade.
\ben
\item
The pion form factors in $\pi \gamma^{\star} \to \gamma$ transition were calculated in
Ref.~\cite{prd76-034008}. The authors found that the NLO correction is only $\sim 5\%$
of the leading order (LO) one when the factorization scale set to be equal with the  momentum transfer.

\item
The pion electromagnetic form factors in $\pi \gamma^{\star} \to \pi$ transition were calculated
in Refs.\cite{prd83-054029,prd89-054015}. The total NLO contribution can provide a roughly
$\sim 20\%$ enhancement to the LO contribution in the considered ranges of the momentum transfer $Q^2$;

\item
The $B \to \pi$ transition form factors involved in the semi-leptonic decay $B \to \pi l \overline{\nu}$
were calculated in Refs.\cite{prd85-074004,prd89-094004}. The NLO contribution from twist-2 part
of the wave function cam provide $\sim 30\%$ correction to the LO order one \cite{prd85-074004},
but it is largely canceled by the
NLO Twist-3 contribution \cite{prd89-094004}, and finally  result in a net $\sim 8\%$ enhancement
to the LO result.

\item
In Ref.~\cite{plb693-102}, the combined analysis of the space-like and time-like electromagnetic pion
form factors has been done in the light-cone pQCD, with the inclusion of the non-perturbative "soft"
QCD and the twist-3 corrections.

\item
The NLO corrections to the time-like pion transition form factor and the
electromagnetic pion form factors have been calculated
in Ref.\cite{plb718-1351}, where the NLO twist-2 correction to the magnitude (phase) are
found to be smaller than $30 \% (30^{\circ})$ for the
time-like pion transition form factors, and lower than $25 \% (10^{\circ})$ for the time-like
electromagnetic form factors at the large invariant mass squared $Q^2>30$ GeV$^2$.
\een
One should note that the vertices for all above mentioned transitions and decay processes
involve the vector currents only. The NLO corrections to the form factors with a scalar vertex, however,
have not been evaluated up to now.

As is well known, the B meson physics is an wonderful place to test the standard model (SM)
and to search for the signal of the new physics (NP) beyond the SM.
For the two body charmless hadronic $B/B_s\to h_1 h_2$ decays (here $h_i$ refers to the light
pseudo-scalar or vector mesons ), such as $B^{\pm} \to \pi^{\pm} \pi^{0}$
and $B^{0} \to \pi^{+} \pi^{-}$ decays, the major contribution come from the factorizable
emission diagrams, in which the space-like form factors with the vector current are  involved.
But for the color-suppressed $B^{0} \to \pi^{0} \pi^{0}$ decay, along with a large
cancelation between the emission diagrams, the contribution from the
annihilation diagrams play an important role: the corresponding amplitude is proportional
to the complex time-like scalar pion form factor.

In the framework of the pQCD factorization approach, the calculations for the main part
of the NLO contributions from various sources have been done during the
past decade for example in Refs.~\cite{prd72-114005,prd85-074004,prd89-094004,plb718-1351},
the still missing NLO part in the pQCD approach includes the ${\cal O}(\alpha^2_s)$ contributions
from the hard spectator diagrams and annihilation diagrams.
In this paper, we will make the first calculation for the scalar pion form factors
up to NLO in the $\kt$ factorization theorem, which has a direct relation with the evaluation of
NLO contributions from those annihilation diagrams of $B/B_s \to h_1 h_2$ decays.

For this purpose,  we firstly give a brief review for the LO space-like and time-like scalar pion
form factors. Secondly, we will calculate the NLO corrections to the space-like scalar pion form
factor, in which the momentum transfer squared is $q^2=-Q^2<0$.
Thirdly, by analytic continuation from $q^2=-Q^2 <0$ to $q^2=Q^2 >0$,
we will obtain the NLO correction to the time-like scalar pion form factors from the one
for the space-like scalar pion form factor.
We finally  evaluate the relevant annihilation diagrams, and to check the effects
on the branching ratios of $B \to \pi\pi$ induced by the inclusion of the newly
found NLO correction to the time-like scalar pion form factors.

By using the universal NLO pion meson wave functions (twist-2 and twist-3 part) as defined in
Refs.~\cite{prd67-034001,prd64-014019,epjc40-395,jhep0601-067},
we will make the convolutions of the LO hard kernel and the NLO wave functions,
evaluate the quark level diagrams for the NLO corrections
to the LO scalar transition process $\pi \to \pi$,
and finally obtain an infrared (IR)  finite NLO hard kernel by making the difference of these two parts.
All the IR singularities, such as the soft divergences generated from exchanging a massless gluon
between two on-shell external quark lines and/or the collinear divergences
generated from emitting a massless gluon from a light parallelled external quark line,
are regulated by the off-shell transverse momentum $k_{\rm iT}$.

We will verify that all the IR divergences obtained from the NLO calculations
can be absorbed into the NLO wave functions completely as described in
Refs.~\cite{prd83-054029,prd89-054015}.
The analytic continuations of the NLO correction to the space-like scalar pion form factor
to the one on the time-like scalar pion form factor is also nontrivial.
When we make the appropriate choice for the renormalization scale $\mu$ and the factorization
scale $\mu_{f}$, say setting them as the internal hard scale $t$ as postulated in
Refs.~\cite{prd83-054029,prd89-054015,plb718-1351},
we find that the NLO corrections on both the space-like and the time-like scalar form factors
are indeed under control naturally.

This paper is organized as follows.
In Sec.~II, we give a brief review about the LO space-like and time-like scalar pion form factors
and show the analytic continuation relation between them.
In Sec.~III, we calculate the NLO correction to the space-like scalar pion form factor and
present the numerical results. In this section, the $O(\alpha^{2}_{s})$ QCD quark diagrams,
as well as the convolutions between LO hard kernel and NLO wave
functions, will be calculated step by step, and then obtain the $\kt$-dependent NLO hard kernel
by making the difference between the two parts.
Sec.~IV contains the analytic continuation of the scalar pion form factor from the space-like domain
$q^2=-Q^2 <0$ to the time-like domain $q^2=Q^2 >0$. We will calculate the branching ratios
for $B \to \pi \pi$ decays, and to check if the ``$\pi\pi$"-puzzle could be
understood by the inclusion of the newly known NLO contribution to the time-like scalar pion
form factor.  Section V finally contains the conclusions.



\section{Leading order analysis}
We first give a brief review about the LO space-like and time-like scalar pion form factors in the framework
of the $\kt$ factorization.
\begin{figure}[tb]
\vspace{-1.5cm}
\begin{center}
\leftline{\epsfxsize=12cm\epsffile{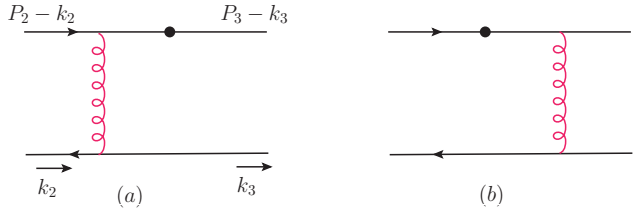}}
\end{center}
\vspace{-13.5cm}
\caption{The LO quark diagrams for the space-like scalar pion form factors for transition $\pi \to \pi$,
with the symbol $\bullet$ representing the insertion of the scalar interaction vertex.}
\label{fig:fig1}
\end{figure}
For the space-like form factor at leading order, the relevant Feynman diagrams are
illustrated in Fig.~\ref{fig:fig1}, where the symbol $\bullet$ representing the insertion of the
scalar interaction vertex. The momentum transfer squared is
\beq
q^{2}=(p_2-p_3)^2=- 2 p_{2} \cdot p_{3} \equiv - Q^{2}, \quad {\rm with} \quad Q^2 > 0,
\eeq
where $p_2$ and $p_3$ are the momentum carried by the initial and final state meson
\beq
p_2=(p^{+}_{2},0,\zerot), \quad p_3=(0,p^{-}_{3},\zerot).
\eeq
While the momentum $k_2$ and $k_3$ carried by the anti-quark in the initial and final state mesons
can be written as
\beq
k_{2} = (x_{2}p^{+}_{2}, 0 , \kta), \quad k_{3} = (0, x_{3}p^{-}_{3}, \ktb),
\eeq
where $x_{2}$ and $x_{3}$ are the momentum fractions of the anti-quark.

\begin{figure}[tb]
\vspace{-1.5cm}
\begin{center}
\leftline{\epsfxsize=12cm\epsffile{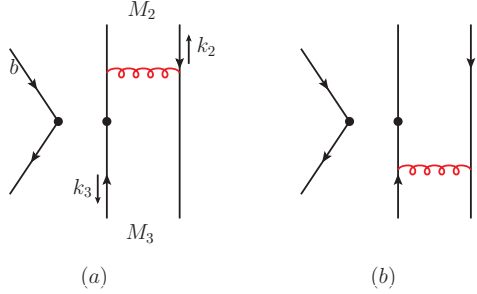}}
\end{center}
\vspace{-12.5cm}
\caption{The LO factorizable annihilation diagrams for  $B \to \pi \pi$ decays,
in which the time-like scalar pion form factors are involved.
The symbol $\bullet$ here represent the effective weak decay vertex.}
\label{fig:fig2}
\end{figure}
For the time-like case as shown in Fig.~\ref{fig:fig2}, the momentum of the initial state B
meson are set as
\beq
p_1=(p^{+}_{1},p^{-}_{1},\zerot ),
\eeq
with the light anti-quark inside the B meson carrying the momentum $k_{1} = (x_{1}p^{+}_{1}, 0 , \kta)$.
The final state mesons $M_2$ and $M_3$, produced from the B meson decay, have the momentum
\beq
p_2=(p^{+}_{2},0,\zerot), \quad p_3=(0,p^{-}_{3},\zerot),
\eeq
and the momentum carried by their partons are defined as
$k_{2}=(x_{2}p^{+}_{2},0,\ktb)$ and $k_{3}=(0,x_{3}p^{-}_{3},\ktc)$ as shown in
Fig.~\ref{fig:fig2} explicitly. But the momentum transfer squared in this process is
$q^2 = (p_2 + p_3)^2 = M^2_B >0$ with $M_B$ being the B meson mass.

Besides the two Feynman diagrams as given in Fig.~\ref{fig:fig1}, in fact, there exist other two Feynman diagrams
with the vertex located in the lower anti-parton lines, which also contribute to  the space-like
scalar pion form factor at the leading order.
But we here just want to evaluate  the NLO corrections to the time-like scalar pion form factors,
which should contribute through the annihilation diagrams of the $B \to M_2 M_3$ decays, as illustrated
in Fig.~\ref{fig:fig2}. For this purpose, only the two Feynman diagrams as shown in Fig.~\ref{fig:fig1}
should be calculated firstly and then we can find the time-like scalar pion form factors
from the space-like ones by making the analytical continuations.
So, we here only consider the two diagrams as given in Fig.~\ref{fig:fig1}.

\subsection{Leading order space-like scalar pion form factor}

Because the vertex in Fig.~\ref{fig:fig1} are scalar in nature, the scalar pion form factors at LO level
can be written directly from the hard kernels of the sub-diagrams in Fig.~\ref{fig:fig1}.
The following hierarchy is postulated in the small-x region in our analytic calculation,
as in Res.~\cite{prd83-054029,prd89-054015}:
\beq
Q^2 \gg x_1 Q^2 \sim x_2 Q^2 \gg x_1 x_2 Q^2 \gg \kta^2 \sim \ktb^2.
\label{eq:hierarchy}
\eeq

\begin{figure}
\centering
\centerline{\epsfxsize=12cm\epsffile{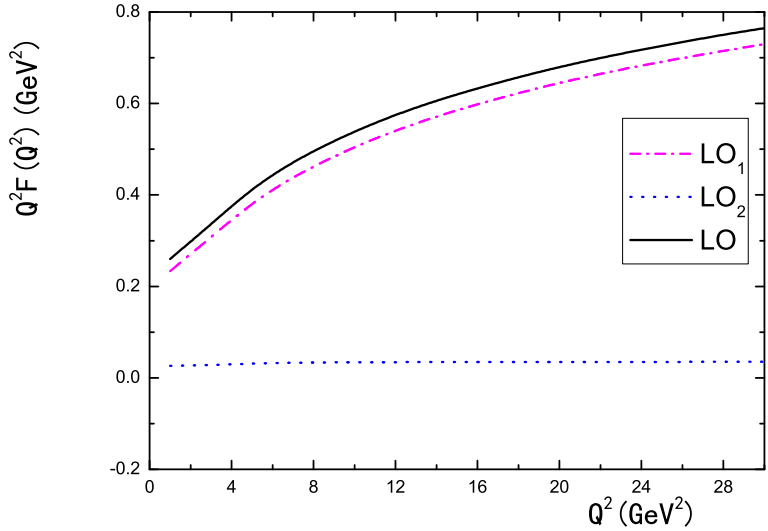}}
\caption{The LO pQCD prediction for the space-like scalar pion form factor
$Q^2F(Q^2)$ from Fig.~\ref{fig:fig1}(a). For more details see the text.}
\label{fig:fig3}
\end{figure}

We use the Fierz identity in Eq.~(\ref{eq:fierz}) and the $SU(3)_c$ group identity in Eq.~(\ref{eq:color})
to factorize the fermion flow and the color flow.
The identity matrix $I$ in the Fierz identity is a $4$-dimension matrix and $(i,j,l,k)$ are the Lorentz index,
while the identity matrix $I$ in the $SU(3)_C$ group is a $3$-dimension matrix and $(i,j,l,k)$ are color index.
\beq
I_{ij}I_{lk} &=& \frac{1}{4}I_{ik}I_{lj} + \frac{1}{4}(\gamma_{5})_{ik}(\gamma_{5})_{lj}
               + \frac{1}{4}(\gamma^{\alpha})_{ik}(\gamma^{\alpha})_{lj}\non
&&               + \frac{1}{4}(\gamma_{5}\gamma^{\alpha})_{ik}(\gamma_{\alpha}\gamma_{5})_{lj}
               + \frac{1}{8}(\sigma^{\alpha\beta}\gamma_{5})_{ik}(\sigma_{\alpha\beta}\gamma_{5})_{lj},
\label{eq:fierz}\\
I_{ij}I_{lk} &=& \frac{1}{N_{c}}I_{lj}I_{ik} + 2(T^{c})_{lj}(T^{c})_{ik}.
\label{eq:color}
\eeq
Because the weak vertex in Fig.~\ref{fig:fig1}(a) is proportional to identity $I$ in the Lorentz space,
then we can obtain the hard kernel $H^{(0)}_{a}$ by sandwiching Fig.~\ref{fig:fig1}(a)
with the following two sets of structures of pion wave functions:
\beq
\left(\frac{\psl_{2}\gamma_{5}}{4N_{c}}, \quad \frac{\gamma_{5}}{4N_{c}}
        ~ or ~ \frac{\gamma_{5}(\nsl_{-}\nsl_{+})}{4N_{c}} \right); ~~~
\left(\frac{\gamma_{5}}{4N_{c}} ~ or ~ \frac{\gamma_{5}(\nsl_{-}\nsl_{+})}{4N_{c}},
        \quad \frac{\gamma_{5}\psl_{3}}{4N_{c}} \right),
\label{eq:sanwich}
\eeq
where $n_{+} = (1,0,\zerot)$ and $n_{-} = (0,1,\zerot)$ denote the unit vector
along with the positive and negative $z$-axis direction.
Of course, one can write down $H^{(0)}_{a}$ directly by using the initial and the final pion meson
wave functions\cite{prd65-014007,zpc48-239,jhep9901-010,jhep0605-004,prd71-014015,csb59-3801} as given
explicitly in Eqs.~(\ref{eq:pi01},\ref{eq:pi02}) with the chiral mass of pion $m_{0\pi}=1.74$ GeV:
\beq
\Phi_\pi(p_2,x_2)&=& \frac{1}{\sqrt{6}} \Bigl \{ \psl_2 \gamma_5 \phi_\pi^A(x_2)
+ m_{0\pi} \gamma_5 \left [ \phi^{P}_{\pi}(x_2) - (\nsl_{-}\nsl_{+} - 1 )\phi^{T}_{\pi}(x_{2}) \right ]
\Bigr \},  \label{eq:pi01}\\
\Phi_\pi(p_3,x_3)&=& \frac{1}{\sqrt{6}}  \Bigl \{ \gamma_5 \psl_3 \phi_\pi^A(x_3)
+ \gamma_5 m_{0\pi} \left [ \phi^{P}_{\pi}(x_3) - ( \nsl_{-}\nsl_{+} - 1 )\phi^{T}_{\pi}(x_3) \right ]
\Bigr\}.  \label{eq:pi02}
\eeq
Then the LO contributions to the hard kernel from Fig.~\ref{fig:fig1}(a) can be written as
\beq
H^{(0)}_{a}(x_{2},x_{3},Q^2) &=& \frac{8 \pi \alpha_s C_F m_{0\pi} Q^2 }{(p_{3}-k_{2})^{2}(k_{2}-k_{3})^{2}}\non
&& \cdot \Bigl  \{ 2\phi^{A}_{\pi}(x_{2}) \phi^{P}_{\pi}(x_{3})
  + x_2 \left [ \phi^{P}_{\pi}(x_{2})-\phi^{T}_{\pi}(x_{2}) \right] \phi^{A}_{\pi}(x_{3})\Bigr \}.
\label{eq:lothka}
\eeq
where $\alpha_s$ is the strong coupling constant, $C_F=4/3$ is the color factor.
It is not difficult to find the end-point behavior of the LO hard kernel for
Fig.~\ref{fig:fig1}(a):
\beq
H^{(0)}_{a}(x_{2},x_{3},Q^2)|_{\rm end-point} &\to& (16 \pi \alpha_s C_F m_{0\pi} Q^2)
       \cdot \Bigl \{\frac{(1-x_2)}{x_2x_3} + (1-x_3) \Bigr\},
\label{eq:endpoint-lothka}
\eeq
where the first and second term describes the end-point behavior  of the corresponding term
in Eq.~(\ref{eq:lothka}). It is easy to verify that the second term in Eq.~(\ref{eq:endpoint-lothka})
is strongly suppressed by a factor of $x_2x_3 (1-x_3)/(1-x_2)$ relative to the first term.
The first term proportioned to $\phi^A(x_3)\phi^P(x_2)$ in Eq.~(\ref{eq:lothka}), consequently,
will provide the dominate contribution when compared with the second term in Eq.~(\ref{eq:lothka}).
The numerical results as illustrated by the curves in Fig.~\ref{fig:fig3} confirmed this point directly.

By using the LO hard kernel $H^{(0)}_a$ in Eq.~(\ref{eq:lothka}),
one can find the corresponding space-like scalar pion form factor at the LO level in the form of
\beq
Q^2F(Q^2)|_{\rm LO} &=& 8 \pi m_{0\pi} C_F Q^4 \int{dx_2 dx_3} \int{b_2 db_2 b_3 db_3} \non
      && \cdot \Bigl \{2 \phi^{A}_{\pi}(x_{2}) \phi^{P}_{\pi}(x_{3}) S_t(x_3)
           + x_2 \left [ \phi^{P}_{\pi}(x_{2}) - \phi^{T}_{\pi}(x_{2}) \right] \phi^{A}_{\pi}(x_{3}) \Bigr\}\non
           && \cdot  \alpha_s(t) \cdot e^{-2 S_{\pi}(t)} \cdot h(x_2,x_3,b_2,b_3),
\label{eq:loff}
\eeq
where the Sudakov factor $S_{\pi}(t)$ and the threshold resummation function $S_t(x)$
are the same ones as being used in Refs.~\cite{prd65-014007,prd89-054015}.
In numerical calculation we choose $c=0.4$ in the function $S_t(x)$.
The hard function $h(x_2,x_3,b_2,b_3)$ in Eq.~(\ref{eq:loff}) can be  written as the following form
\beq
h(x_2,x_3,b_2,b_3)&=&K_0\left (\sqrt{x_2x_3} Q b_2 \right)\non
 &&\cdot   \Bigl [\theta(b_2-b_3)I_0\left (\sqrt{x_3} Q b_3 \right) K_0\left (\sqrt{x_3} Q b_2 \right)
          + (b_2 \leftrightarrow b_3)\Bigr],
\eeq
where the function $K_0$ and $I_0$ are the modified Bessel function.
Following Refs.~\cite{prd83-054029,prd89-054015}, we here also choose $\mu=\mu_f=t$ in the numerical calculations:
\beq
\mu=\mu_f= t = \max\left (\sqrt{x_2} Q, \sqrt{x_3}Q, 1/b_2,1/b_3 \right).
\label{eq:mumuf}
\eeq

In the calculations, we can consider the sub-diagram Fig.~\ref{fig:fig1}(a) only,
since the contributions from Fig.~\ref{fig:fig1}(b) can be obtained by simple kinematic replacements
of $x_2\leftrightarrow x_3$ for the results from the Fig.~\ref{fig:fig1}(a)
as we have argued in Ref.~\cite{prd89-054015}.
The direct analytical evaluations for Fig.~\ref{fig:fig1}(b) can verify this exchange symmetry.
After making the analytic calculations  we found the expressions for the LO hard kernel $H^{(0)}_{b}(x_{1},x_{2},Q^2)$ and its
end-point behavior:
\beq
H^{(0)}_{b}(x_{2},x_{3},Q^2) &=& \frac{8 \pi \alpha_s C_F m_{0\pi} Q^2 }{(p_{2}-k_{3})^{2}(k_{3}-k_{2})^{2}}\non
&& \cdot \Bigl  \{ 2\phi^{A}_{\pi}(x_{3}) \phi^{P}_{\pi}(x_{2})
  + x_3 \left [ \phi^{P}_{\pi}(x_{3})-\phi^{T}_{\pi}(x_{3}) \right] \phi^{A}_{\pi}(x_{2})\Bigr \},\ \  \label{eq:lothkb} \\
H^{(0)}_{b}(x_{2},x_{3},Q^2)|_{\rm end-point} &\rightarrow& (16 \pi \alpha_s C_F m_{0\pi} Q^2)
       \cdot \Bigl \{\frac{(1-x_3)}{x_2x_3} + (1-x_2) \Bigr\}.
\label{eq:endpoint-lothkb}
\eeq

In Fig.~\ref{fig:fig3}, we show the $Q^2$-dependence of the LO space-like scalar pion form factor
$Q^2 F(Q^2)$ for Fig.~\ref{fig:fig1}(a), in order to support our previous theoretical arguments for the
dominance of the contribution from the first term of the hard kernel $H^{(0)}_{a}(x_{2},x_{3},Q^2)$ as
defined in Eq.~(\ref{eq:lothka}).
In Fig.~\ref{fig:fig3}, the contributions from the two different terms
as given in Eq.~(\ref{eq:lothka}) are plotted explicitly:
the upper dot-dashed curve with the label "LO$_1$" shows the contribution from the first term
proportional to $2\phi^{P}_{\pi}(x_{2}) \phi^{A}_{\pi}(x_{3})$ in the LO hard kernel $H^{(0)}_{a}$,
while the lower doted curve with the label "LO$_2$" shows the contribution from the second
term in $H^{(0)}_{a}$, and finally the solid line denotes the total LO contribution.
One can see from the curves in Fig.~\ref{fig:fig3} that the contribution to the LO space-like
scalar pion form factor $Q^2 F(Q^2)$ from the first term of  $H^{(0)}_{a}$ is
indeed dominant absolutely, larger than $90\%$ of the total LO result
in the whole considered range of $Q^2$.

In the numerical calculations, we integrate for the partons' momentum fractions ($x_2,x_3$) over the
range of $x_i=[0,1]$, and find that the main contribution comes from the small
$x_2,x_3 \sim 0.1$ region \cite{prd66-094010}, being consistent with the hierarchy postulated
in Eq.~(\ref{eq:hierarchy}) for our analytic calculation.
After the numerical integration, we obtained the LO theoretical predictions as shown in Fig.~\ref{fig:fig3}
by using the the ordinary full pion distribution amplitudes(DA's) as given in
Refs.~\cite{prd71-014015,jhep0605-004}:
\beq
\phi_{\pi}^{A}(x) &= &\frac{3 f_{\pi}}{\sqrt{6}} x (1-x)
\left [ 1 +  a_2^{\pi} C_2^{\frac{3}{2}}(u) + a_4^{\pi} C_4^{\frac{3}{2}}(u)
                    \right ], \non
\phi_{\pi}^{P}(x) &= &\frac{f_{\pi}}{2\sqrt{6}} \left[ 1 + \left (30 \eta
_3 - \frac{5}{2} \rho_{\pi}^2 \right ) C_2^{\frac{1}{2}}(u) - 3 \left (\eta_3 \omega_3
+ \frac{9}{20} \rho_{\pi}^2 \left (1 + 6 a_2^{\pi}\right ) \right ) C_4^{\frac{1}{2}}(u) \right ],  \non
\phi_{\pi}^{T}(x) &= & \frac{f_{\pi}}{2\sqrt{6}} (1-2x)
\left [1 + 6 \left (5 \eta_3 - \frac{1}{2} \eta_3 \omega_3 - \frac{7}{20} \rho_{\pi}^2
- \frac{3}{5} \rho_{\pi}^2 a_2^{\pi} \right ) \left (1-10 x + 10 x^2 \right )  \right ],\non
\label{eq:phipi1}
\eeq
where the pion decay constant $f_\pi=0.13$ GeV, the Gegenbauer moments $a_i^\pi$, the parameters
$\eta_3, \omega_3$ and $\rho_\pi$ are adapted from Refs.~\cite{jhep0605-004,prd71-014015}:
\beq
a_2^{\pi} &=& 0.25\pm 0.015, \quad a_4^{\pi} = -0.015, \quad
\rho_{\pi} = m_{\pi}/m^0_{\pi}, \quad \eta_3 = 0.015, \quad \omega_3 = -3.0.
\label{eq:input1}
\eeq
The relevant Gegenbauer polynomials $C_{2,4}^{1/2}(2x-1)$ and $C_{2,4}^{3/2}(2x-1)$
can be found easily in Refs.~\cite{jhep0605-004,prd71-014015}.

From the expressions of the LO hard kernels $H^{(0)}_a,H^{(0)}_b$ as given in Eqs.~(\ref{eq:lothka},\ref{eq:lothkb})
and the LO pQCD predictions for $Q^2 F(Q^2)$ as illustrated in Fig.~\ref{fig:fig3}, one can see the following points:
\begin{enumerate}
\item
The LO hard kernel $H^{(0)}$ only receive the contributions from the two cross productions of the DA's with different
twists for the initial and final pion wave function, because of the nature of the scalar current in the vertex of
Fig.~\ref{fig:fig1}.
Take $H^{(0)}_a$ as an example, the contributions do come from the two terms proportional to
$\phi^{A}(x_{2}) \phi^{P}(x_{3})$ and $[\phi^P(x_2) - \phi^T(x_2)]\phi^A(x_3)$ respectively, as listed
in Eq.~(\ref{eq:lothka}).

\item
From Fig.~\ref{fig:fig3} one can see easily that the first term in the LO hard kernel $H_{a}^{(0)}$
in Eq.~(\ref{eq:lothka}) provides the absolutely
dominant contribution (larger than $90\%$) to the form factor $Q^2F(Q^2)$ in the whole
region of $1< Q^2 < 30$ GeV$^2$,
while the contribution from the second term of $H_{a}^{(0)}$ in Eq.~(\ref{eq:lothka}) is very small
and can be neglected safely.
This fact does support our previous argument from the analysis for the end-point
behavior of the two terms
in Eqs.~(\ref{eq:endpoint-lothka},\ref{eq:endpoint-lothkb}).

\item
Since the LO contribution from the second term of $H^{(0)}_a$ in Eq.~(\ref{eq:lothka})
is already very small,
it is reasonable for us to consider the NLO contribution to the space-like scalar pion form factor
from the dominant first term in $H^{(0)}_{a}$ only in next section,
which would simplify our calculations significantly.
\end{enumerate}

\subsection{Leading order time-like scalar pion form factor}

The weak decay vertices in the factorizable annihilation diagrams
for B decays, as shown in Fig.~\ref{fig:fig2},
would generate three kinds of contributions: $(V-A)\otimes(V-A)$,
$(V-A)\otimes(V+A)$ and $(S-P)\otimes(S+P)$ current contribution.
We here abbreviate these contributions as LL, LR and SP current respectively.
The SP current comes from the Fierz transformation of the LR current if the light anti-quarks in B meson
and one of the final light mesons are identical.
In this paper, we will just consider the SP current proportioned
to the time-like scalar pion form factors,
because the LO hard kernels with the LL and LR currents are canceled each other completely
in Fig.~\ref{fig:fig2}(a) and Fig.~\ref{fig:fig2}(b) when the two
final states have the same wave functions,
as described in detail in Ref.~\cite{prd63-074009}.

Using the definition of the annihilation matrix element $<0|(\bar{q}b)_{S-P}|B(p_1)>=if_B M_B$, the LO hard kernel
for Fig.~\ref{fig:fig2}(a) and \ref{fig:fig2}(b) can be written in the following form:
\beq
H'^{(0)}_{a}(x_{2},k_{\rm 2T};x_{3},k_{\rm 3T};M^2_B)&=&
\frac{-8\pi \alpha_s C_F m^{0}_{\pi} M^2_B (-2 i f_B M_B)}
{[(p_{1}-k_{3})^{2}+i\epsilon] ](k_{3}-p_{3}-k_{2})^{2}+i\epsilon]} \non
 && \hspace{-2cm}\cdot
 \Bigl \{ (1 - x_3) \phi^{A}_{\pi}(x_{2}) \left [ \phi^{P}_{\pi}(x_{3}) + \phi^{T}_{\pi}(x_{3}) \right]
       + 2 \phi^{P}_{\pi}(x_{2}) \phi^{A}_{\pi}(x_{3}) \Bigr \}, \label{eq:lohka} \\
H'^{(0)}_{b}(x_{2},k_{\rm 2T};x_{3},k_{\rm 3T};M^2_B)&=&
\frac{-8\pi \alpha_s C_F m^{0}_{\pi} M^2_B (-2 i f_B M_B)}
{[(k_{3}-p_{3}-k_{2})^{2}+i\epsilon] [(p_{3}+k_{2})^{2}+i\epsilon]} \non
&&\hspace{-2cm}\cdot \Bigl \{x_2 \left[ \phi^{P}_{\pi}(x_{2}) - \phi^{T}_{\pi}(x_{2}) \right]
\phi^{A}_{\pi}(x_{3})
         + 2 \phi^{A}_{\pi}(x_{2}) \phi^{P}_{\pi}(x_{3}) \Bigr \}. \label{eq:lohkb}
\eeq
It's easy to confirm that the LO hard amplitude $H'^{(0)}_{b}$ can be obtained from $H'^{(0)}_{a}$  by
simple kinetic replacements $x_3 \leftrightarrow (1-x_2)$, we therefore could deal with the
Fig.~\ref{fig:fig2}(a) in our calculation and get the one for Fig.~\ref{fig:fig2}(b) by proper
kinetic replacements $x_3 \leftrightarrow (1-x_2)$.

Furthermore, the contribution to $H'^{(0)}_{a}$ mainly comes  from the small $(1-x_3) \sim 0.1$ region
due to the threshold suppression effects\cite{prd66-094010},
while the contribution proportional to term $\phi^{A}_{\pi}(x_{2})
\left [\phi^{P}_{\pi}(x_{3}) + \phi^{T}_{\pi}(x_{3} \right]$
in $H'^{(0)}_{a}$ is also suppressed by the factor $x_2(1-x_2)(1-x_3)/x_3$.
It is therefore reasonable for us to consider the dominant
$H'^{(0)}_{a,32}$ proportioned to term $\phi^{P}_{\pi}(x_{2}) \phi^{A}_{\pi}(x_{3})$
only when we calculate the hard amplitude $H'^{(0)}_{a}$:
\beq
H'^{(0)}_{a,32}(x_{2},k_{\rm 2T};x_{3},k_{\rm 3T};M^2_B) \equiv
   \frac{-16 \pi \alpha_s C_F m^{0}_{\pi} M^2_B (-2 i f_B M_B)}
   {[(p_{1}-k_{3})^{2}+i\epsilon] [(k_{3}-p_{3}-k_{2})^{2}+i\epsilon]}
    \phi^{P}_{\pi}(x_{2}) \phi^{A}_{\pi}(x_{3}). \ \
\label{eq:lohka32}
\eeq

For the Fourier transformation of Eq.~(\ref{eq:lohka}) from the transverse-momentum
space $(\ktb, \ktc)$ to the impact-parameter space $(b_2, b_3)$, we have two different
choices: One is the double-b convolution, another is the single-b convolution.

If one write the two factors in the denominator of the hard
kernel in Eqs.~(\ref{eq:lohka},\ref{eq:lohka32}) in the form of
\beq
(p_{1}-k_{3})^{2}+i\epsilon &=& M^2_B (1-x_3) - \ktc^2 +i\epsilon; \non
(k_{3}-p_{3}-k_{2})^{2}+i\epsilon &=& M^2_B x_2 (1-x_3) - (\ktc - \ktb)^2 +i\epsilon,
\label{eq:denominatorII}
\eeq
one could obtain, after making the integration of the hard kernel in Eq.~(\ref{eq:lohka})
over the whole momentum space, the complex time-like amplitudes which may provide a strong phase to
generate the large CP violation observed for example in $B \to \pi^+\pi^-$ decay \cite{prd63-074009}.
The imaginary part of the resulted time-like amplitude is produced according to the
principle-value prescription
in Eq.~(\ref{eq:principle-value}) when one of the internal particle propagators goes on mass-shell:
\beq
\frac{1}{x M^2_B - \kt^2 \pm i\epsilon} = {\rm Pr} \left(\frac{1}{x M^2_B - \kt^2} \right) \mp i \pi \delta(x M^2_B - \kt^2).
\label{eq:principle-value}
\eeq
Then the LO time-like scalar pion form factor for Fig.~\ref{fig:fig2}(a) can be obtained
by the Fourier transformation of Eq.~(\ref{eq:lohka})
from the space $(\ktb, \ktc)$ to the space $(b_2, b_3)$,
and this double-b convolution can then be written in the form of
\beq
F'^{(0)}_{a,II} &=& \int^1_0 dx_2 dx_3 \int^{\infty}_0 db_2 db_3 ~ 16 \pi C_F M^4_B m^{\pi}_0
                   \cdot \alpha_s(\mu) \cdot
\exp\left [-S_{\rm II}(x_2,b_2;1-x_3,b_3;M_B;\mu) \right ] \non
&& \hspace{-1cm}\cdot \Bigl \{ (1 - x_3) \phi^A_{\pi}(x_2)\left [ \phi^P_{\pi}(x_3) + \phi^T_{\pi}(x_3) \right]
                   + 2 \phi^P_{\pi}(x_2)\phi^A_{\pi}(x_3) \cdot S_{t}(x_2) \Bigr \} \non
                && \hspace{-1cm}\cdot K_0(i \sqrt{(1-x_3)x_2} M_B b_3)
                   \cdot \Bigl [ K_0(\sqrt{x_2} M_B b_2) I_0(\sqrt{x_2} M_B b_3) \theta(b_2-b_3)
                             + (b_2 \leftrightarrow b_3) \Bigr ].\ \
\label{eq:loffII}
\eeq

If one take the hierarchy as described in Eq.~(\ref{eq:hierarchy}) into account, he can also
drop the transverse momentum of the internal quark but keep the transverse momentum of the gluon
propagator, i.e. write the two factors of the denominator in Eqs.~(\ref{eq:lohka},\ref{eq:lohka32})
in the form of
\beq
(p_{1}-k_{3})^{2}+i\epsilon &\sim& M^2_B (1-x_3) + i \epsilon,  \non
(k_{3}-p_{3}-k_{2})^{2}+i\epsilon &=& M^2_B x_2 (1-x_3) - (\ktc - \ktb)^2 +i\epsilon,
\label{eq:denominatorI}
\eeq
One can then obtain the single-b convolution LO time-like scalar pion form factor for
Fig.~\ref{fig:fig2}(a) by Fourier transformation of the Eq.~(\ref{eq:lohka}) from the
transverse-momentum space $\ktc$ to the impact-parameter space $b_3$
with only one b parameter integration.
This single-b convolution form factor is in the form of
\beq
F'^{(0)}_{a,I} =&& \int^1_0 dx_2 dx_3 \int^{\infty}_0 db_3 ~ \frac{- 4 C_F M^2_B m^{\pi}_0}{(1-x_3)}
                   \cdot \alpha_s(\mu) \cdot \exp\left [-S_{\rm I}(x_2,b_2;1-x_3,b_3;M_B;\mu) \right] \non
               && \cdot \Bigl \{ (1 - x_3) \phi^A_{\pi}(x_2)
\left [\phi^P_{\pi}(x_3) + \phi^T_{\pi}(x_3) \right]
+ 2 \phi^P_{\pi}(x_2)\phi^A_{\pi}(x_3) \cdot S_{t}(x_2)\Bigr \} \non
               &&  \cdot K_0(i \sqrt{(1-x_3)x_2} M_B b_3).
\label{eq:loffI}
\eeq
The Bessel functions $K_0, I_0$ and the Sudakov exponents  $S_{I,II}$
in Eqs.~(\ref{eq:loffII},\ref{eq:loffI}) are of the form
\beq
K_0(iz)&=&\frac{i\pi}{2} H^{(1)}_0(iz) = \frac{i\pi}{2}\left
[ I_0(z)+i N_0(z)\right], \label{eq:Bessel} \\
S_{\rm I}&=& S_{\rm II}=S(x_2,b_2;M_B;\mu)+S(1-x_3,b_3;M_B;\mu). \label{eq:Sudakov}
\eeq

\begin{figure*}
\centering
\vspace{0cm}
\includegraphics[width=0.51\textwidth]{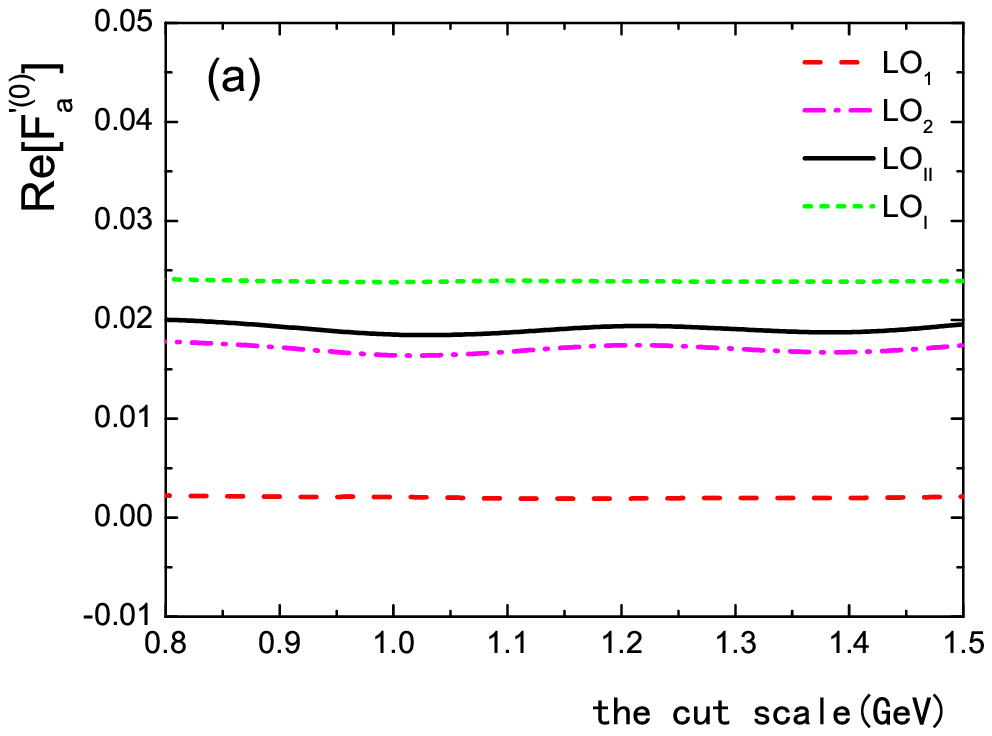}
\hspace{-1.2cm}
\includegraphics[width=0.51\textwidth]{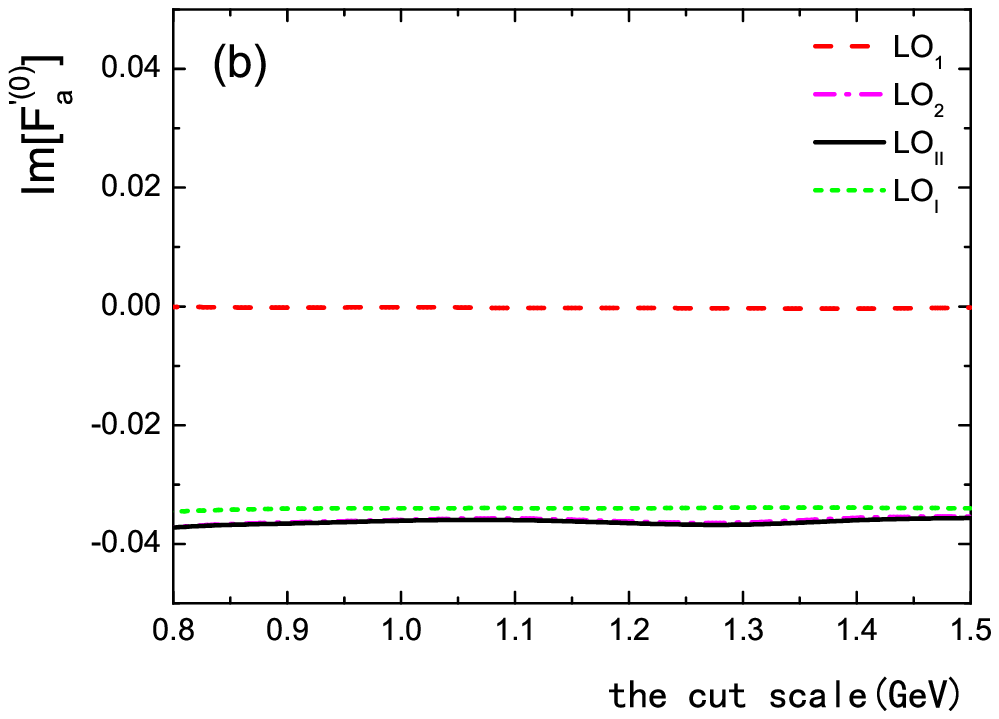}
\vspace{0cm}
\caption{The $\mu_0$-dependence of the LO time-like scalar pion form factor $F'^{(0)}_{a,I} $  and $F'^{(0)}_{a,II} $
on the  variations of the cutoff scale $\mu_0$ in the range of $0.8 \leq \mu_0\leq 1.5$ GeV.
(a) the real part ${\rm Re}[F'^{(0)}_{a}]$, and (b) the imaginary part ${\rm Im}[F'^{(0)}_{a}]$. For more details see text. }
\label{fig:fig4}
\end{figure*}

In the framework of the pQCD factorization approach, we usually choose
the lower cutoff scale $\mu_0=1$ GeV for the hard scale t in the running of the Wilson coefficients
$C_i(t)$. In the numerical integrations we will fix the values $C_i(t)$ at $C_i(\mu_0)$ whenever
the scale $t$ runs below the cutoff scale $\mu_0=1$ GeV.

In Fig.~\ref{fig:fig4}, we plot the pQCD predictions for the values and $\mu_0$-dependence of the real and imaginary
part of the LO time-like scalar pion form factor  $F'^{(0)}_{a,I} $  and $F'^{(0)}_{a,II} $,
obtained by using the pion DAs as given in Eq.~(\ref{eq:phipi1}) and the single-b and double-b
convolution respectively.
The Fig.~4(a) and  4(b) shows the real and the imaginary part of the form factor, respectively.
The dash, dot-dash curve represents the contribution from the 1st, 2nd-term in Eq.~(\ref{eq:loffII}) respectively;
The solid ( dots) line shows the total LO form factor as described in Eq.~(\ref{eq:loffII}) ( Eq.~(\ref{eq:loffI})),
when making the double-b (single-b) convolution.
From the numerical results as shown in Fig.~4, we find the following points:
\begin{enumerate}
\item
The second term proportional to $\phi^{P}_{\pi}(x_{2}) \phi^{A}_{\pi}(x_{3})$ in Eq.~(\ref{eq:loffII}) provides
the dominant contribution to both the real and imaginary part of the time-like scalar pion
form factor, which support our analysis in the paragraph before Eq.~(\ref{eq:lohka32})
and further imply that it's reasonable to consider the NLO contribution to this dominant term only
when we evaluate the NLO corrections to the LO results.

\item
The real (imaginary) part of the form factor obtained from the single-b convolution
is a bit larger (smaller) than that obtained from the double-b convolution method.
But one can see from Fig.~4 that the difference between the pQCD predictions obtained by using the single-b or double-b
convolution method is very small in deed.
Such similarity  can be understood by  the fact that the internal gluon propagator provide the major contribution to
the imaginary part and the strong phase in the factorizable annihilation diagrams.

\item
It is easy to see from Fig.~4 that the pQCD predictions for the LO time-like scalar pion form factor
$F'^{(0)}_{a,I} $  and $F'^{(0)}_{a,II} $ do have a very weak dependence on the  value of $\mu_0$,
this is indeed what we expect.

\end{enumerate}

By comparing the hard kernels as given in Eqs.~(\ref{eq:lothka},\ref{eq:lohka}),
it is easy to find that one can obtain the time-like hard kernel $H'^{(0)}_a$ from the
space-like one $H^{(0)}_a$
by simple replacements of $x_2 \rightarrow 1-x_3$ and the analytic continuation $Q^2 \rightarrow -M^2_B$.
Such connections are also valid for $H'^{(0)}_b$ and $H^{(0)}_b$.
So the NLO contribution to the LO time-like scalar hard kernel can also be obtained from the
NLO correction to the LO space-like result by the same kinds of replacements and analytical
continuations, which will be presented in the next section.

\section{NLO correction for the space-like scalar pion form factor}

In this section we will calculate the ${\calo}(\alpha^2_s)$ quark level diagrams as well as the convolutions of the effective diagrams for the ${\calo}(\alpha_s)$ wave functions and the LO (${\calo}(\alpha_s)$) hard kernel in the 't Hooft-Feynman gauge,
and try to find the IR finite NLO corrections to  the space-like scalar pion form factor in the $\kt$ factorization theorem.
From the discussions in last section, we get to know that it is reasonable for us
to calculate the NLO corrections to $H^{(0)}_{a,1}$ only, which is the the
first term of the LO hard kernel $H^{(0)}_{a}(x_{2},x_{3},Q^2)$ in Eq.~(\ref{eq:lothka}), i.e.,
\beq
H^{(0)}_{a,1} (x_{2},k_{\rm 2T};x_{3},k_{\rm 3T};Q^2)=
   \frac{16 \pi \alpha_s C_F m_{0\pi} Q^2}{(p_{3}-k_{2})^{2}(k_{2}-k_{3})^{2}}
   \; \phi^{A}_{\pi}(x_{2}) \phi^{P}_{\pi}(x_{3}).
\label{eq:lothka1}
\eeq
Under the hierarchy as shown in Eq.~(\ref{eq:hierarchy}), only those terms which don't vanish in
the limits of $x_i \to 0$ and $k_{iT} \to 0$ should be kept.

\subsection{NLO contributions of the QCD quark diagrams}

\begin{figure}[tb]
  \centering
  \vspace{-1cm}
  \begin{center}
  \leftline{\epsfxsize=10cm\epsffile{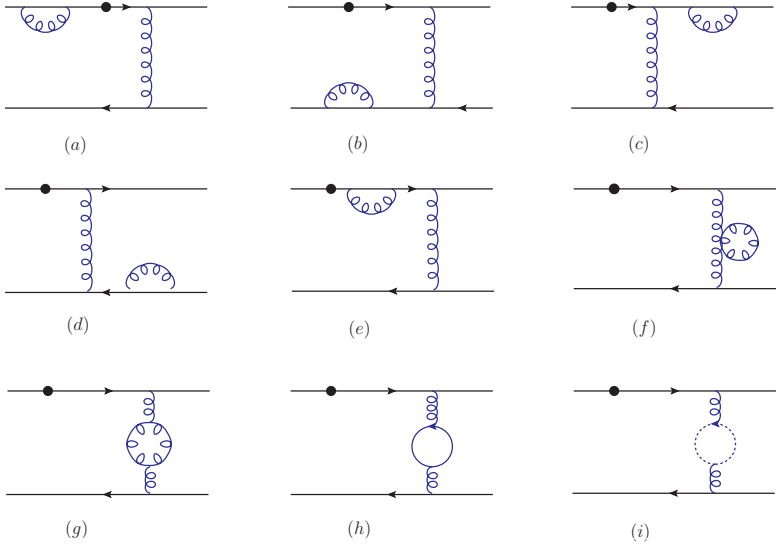}}
  \end{center}
  \vspace{-6.5cm}
  \caption{The self-energy corrections to Fig.1(a).}
  \label{fig:fig5}
  \end{figure}

\begin{figure}[tb]
\vspace{-1cm}
\begin{center}
\leftline{\epsfxsize=10cm\epsffile{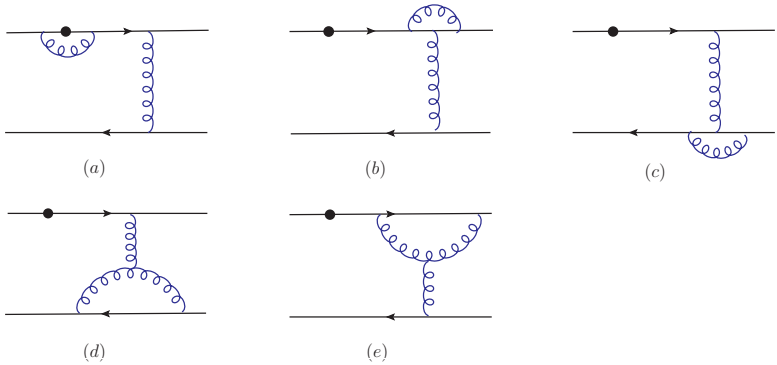}}
\end{center}
\vspace{-9cm}
\caption{The vertex corrections to Fig.1(a).}
\label{fig:fig6}
\end{figure}

\begin{figure}[tb]
\vspace{-1cm}
\begin{center}
\leftline{\epsfxsize=10cm\epsffile{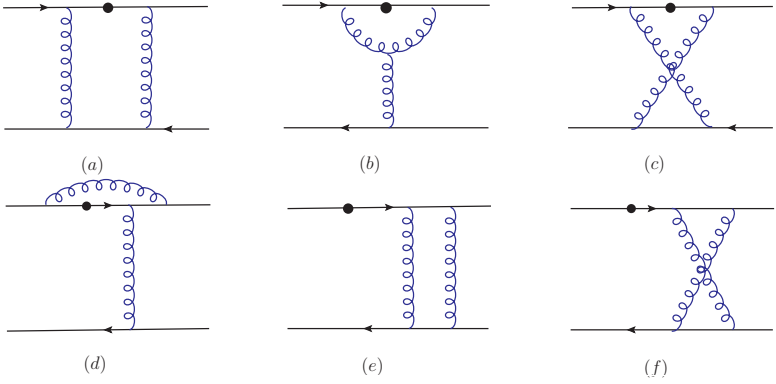}}
\end{center}
\vspace{-9cm}
\caption{The Box and pentagon corrections to Fig.1(a).}
\label{fig:fig7}
\end{figure}

We first calculate the NLO (${\calo}(\alpha^2_s)$) corrections to Fig.~\ref{fig:fig1}(a) in the $\kt$ factorization theorem in this subsection.
These NLO corrections include the self-energy diagrams, the vertex diagrams, the box and pentagon diagrams,
as illustrated in Figs.~(\ref{fig:fig5},\ref{fig:fig6},\ref{fig:fig7}) respectively.
We will use the dimensional reduction scheme\cite{plb84-193} to extract the ultraviolet (UV) divergences,
and use the transverse momentum for the external light quarks in Eq.~(\ref{eq:defi1}) to regulate the IR divergences in loops.
Following the method used in Refs.~\cite{prd83-054029,prd89-054015} we make the same
definitions for $\delta_2, \delta_3$ and $\delta_{23}$:
\beq
\delta_2 = \frac{k^2_{2T}}{Q^2}, \quad
\delta_3 = \frac{k^2_{3T}}{Q^2}, \quad
\delta_{23} = \frac{-(k_2 - k_3)^2}{Q^2}.
\label{eq:defi1}
\eeq

Following the standard procedure we calculate the one loop self-energy Feynman diagrams
as shown in Fig.~\ref{fig:fig5} and find the
following NLO self-enegy corrections:
\beq
&&G^{(1)}_{5a} = -\frac{\alpha_s C_F}{8 \pi}\left [\frac{1}{\epsilon} +
\ln{\frac{4 \pi \mu^2}{\delta_2 Q^2 e^{\gamma_E}}} + 2 \right ]H^{(0)},\non
&&G^{(1)}_{5b} = -\frac{\alpha_s C_F}{8 \pi}\left [\frac{1}{\epsilon} +
\ln{\frac{4 \pi \mu^2}{\delta_2 Q^2 e^{\gamma_E}}} + 2 \right ]H^{(0)},\non
&&G^{(1)}_{5c} = -\frac{\alpha_s C_F}{8 \pi}\left [\frac{1}{\epsilon} +
\ln{\frac{4 \pi \mu^2}{\delta_3 Q^2 e^{\gamma_E}}} + 2 \right ]H^{(0)},\non
&&G^{(1)}_{5d} = -\frac{\alpha_s C_F}{8 \pi}\left [\frac{1}{\epsilon} +
\ln{\frac{4 \pi \mu^2}{\delta_3 Q^2 e^{\gamma_E}}} + 2 \right ]H^{(0)},\non
&&G^{(1)}_{5e} = -\frac{\alpha_s C_F}{4 \pi} \left [\frac{1}{\epsilon} +
\ln{\frac{4 \pi \mu^2}{x_2 Q^2 e^{\gamma_E}}} + 2 \right ]H^{(0)},\non
&&G^{(1)}_{5f+5g+5h+5i} = \frac{\alpha_s}{4 \pi} \left [\left (5 -\frac{2}{3} N_f \right )
\left  (\frac{1}{\epsilon} + \ln{\frac{4 \pi \mu^2}{\delta_{23}Q^2 e^{\gamma_E}}} \right ) \right ]H^{(0)},
\label{eq:self}
\eeq
where $1 / \epsilon$ represents the UV pole term, $\mu$ is the renormalization scale,$\gamma_E$ is the Euler constant,
$N_f$ is the number of the active quarks flavors, and $H^{(0)}= H^{(0)}_{a,1} ( x_{2},k_{\rm 2T};x_{3},k_{\rm 3T};Q^2)$
has been defined in Eq.~(\ref{eq:lothka1}).
For the sake of simplicity, we will use the abbreviation $H^{(0)}$ instead
of the term $H^{(0)}_{a,1} ( x_{2},k_{\rm 2T};x_{3},k_{\rm 3T};Q^2)$ to denote the LO hard kernel throughout the text
unless otherwise stated explicitly.
The Figs.~\ref{fig:fig5}(f,g,h,i) denote the self-energy corrections to the exchanged gluon itself.

It's easy to find that all these self energy corrections are equal to the self energy corrections for the pion
electromagnetic form factors\cite{prd83-054029,prd89-054015}, because these self energy diagrams just correct
the light quark fields, while don't involve the inner structure of the initial and final mesons.
The additional factor $1/2$ is considered for self energy diagrams Fig.~\ref{fig:fig5}(a,b,c,d) because of the freedom to choose
the most outside vertex of the radiative gluon.

The vertex correction diagrams with three-point loop integrals are plotted in Fig.~\ref{fig:fig6},
the NLO corrections from these five vertex diagrams are summarized in the following form:
\beq
G^{(1)}_{6a} &=&  \frac{\alpha_s C_F}{4 \pi} \left [\frac{1}{\epsilon} +
       \ln{\frac{4 \pi \mu^2}{Q^2 e^{\gamma_E}}} - 2 \ln{x_2}\ln{\delta_2}
       - 2 \ln{\delta_2} - 2 \ln{x_2} - \frac{5 \pi^2}{2} +2 \right ]H^{(0)}, \non
G^{(1)}_{6b} &=& -\frac{\alpha_s}{8 \pi N_c} \left [\frac{1}{\epsilon} +
       \ln{\frac{4 \pi \mu^2}{x_2 Q^2 e^{\gamma_E}}} + 1 \right ]H^{(0)}, \non
G^{(1)}_{6c} &=& -\frac{\alpha_s}{8 \pi N_c} \left [\frac{1}{\epsilon} +
       \ln{\frac{4 \pi \mu^2}{\delta_{23} Q^2 e^{\gamma_E}}}
       - \ln{\frac{\delta_2}{\delta_{23}}} \ln{\frac{\delta_3}{\delta_{23}}}
       - \ln{\frac{\delta_3}{\delta_{23}}} - \ln{\frac{\delta_3}{\delta_{23}}} - \frac{\pi^2}{3} +2 \right ]H^{(0)}, \non
G^{(1)}_{6d} &=& \frac{\alpha_s N_c}{8 \pi} \left [\frac{3}{\epsilon} +
       3 \ln{\frac{4 \pi \mu^2}{\delta_{23} Q^2 e^{\gamma_E}}}
       - \ln{\frac{\delta_2}{\delta_{23}}} - \ln{\frac{\delta_3}{\delta_{23}}} +6 \right ]H^{(0)}, \non
G^{(1)}_{6e} &=& \frac{\alpha_s N_c}{8 \pi} \left [\frac{3}{\epsilon} +
       3 \ln{\frac{4 \pi \mu^2}{x_2 Q^2 e^{\gamma_E}}} -  \ln{x_3}\ln{\delta_3} - \ln{\delta_3}\right. \non
 &&\left.       + \ln{x_2}\ln{x_3} + \ln{x_3} - \frac{\pi^2}{3} +6 \right ]H^{(0)}.
\label{eq:vertex}
\eeq
All these five vertex diagrams would have IR divergences at the first sight.
The radiated gluon in Fig.~\ref{fig:fig6}(a) would generate the collinear divergence when it's parallel to the initial momentum $p_2$.
Fig.~\ref{fig:fig6}(b) would include the collinear divergence at $l \parallel p_3$ region.
Fig.~\ref{fig:fig6}(c) would generate both the soft and collinear divergence because the radiated gluon is attached to the external light quark lines, then the double logarithm would appear.
The radiated gluon in Fig.~\ref{fig:fig6}(d) would generate the collinear divergences from $l \parallel p_2$ and $l \parallel p_3$ regions,
while the gluon in Fig.~\ref{fig:fig6}(e) could generate the collinear divergence in the $l \parallel p_2$ region.
But the detailed calculations show that the collinear singularity in Fig.~\ref{fig:fig6}(b) is forbidden by the kinetics,
so $G^{(1)}_{6b}$ is IR finite.

The box and pentagon diagram in Fig.~\ref{fig:fig7} are more complicate because they
would involve four-point and five-point integrals. But the sub-diagrams Figs.~\ref{fig:fig7}(a,e)
are reducible diagrams and their contributions
will be canceled completely by the relevant effective diagrams to be evaluated in the next
subsection, so we can set them to be zero here safely.
Then we just need to calculate three four-point diagrams Figs.~\ref{fig:fig7}(b,d,f) and one five-point
diagram Fig.~\ref{fig:fig7}(c).  From the evaluations of
the Feynman diagrams in Fig.~\ref{fig:fig7} we find the following NLO corrections:
\beq
G^{(1)}_{7a,7e} &\equiv &0, \non
G^{(1)}_{7b} &=& - \frac{\alpha_s  N_c}{8 \pi}
                 \left [ \ln{\delta_{2}} - \ln{\delta_{23}} -1 \right ]H^{(0)}, \non
G^{(1)}_{7c} &=& - \frac{\alpha_s}{8 \pi  N_c}
                 \left [ \ln{\delta_{2}} \ln{\delta_{3}} - 2 \ln{x_2} \ln{\delta_2} - \ln{\delta_2}
                 + \frac{1}{2} \ln^2{\delta_{23}} - \ln^{2}{x_{3}} - \frac{5}{12} \pi^2 \right ]H^{(0)},\non
G^{(1)}_{7d} &=&   \frac{\alpha_s}{8 \pi N_c}
                 \left [ \ln{\delta_{2}} \ln{\delta_{3}} - 2 \ln{x_2} \ln{\delta_2} + \ln{\delta_3}
                 - \ln{x_{2}} - \frac{\pi^2}{3} -1 \right ] H^{(0)}, \non
G^{(1)}_{7f} &=& - \frac{\alpha_s}{8 \pi N_c}
                 \left [ \ln{\frac{\delta_1}{\delta_{23}}} \ln{\frac{\delta_3}{\delta_{23}}}
                 - \ln{x_3} \ln{\delta_3} + \frac{1}{2} \ln^2{\delta_{23}} \right.\non
&&\left.                  + \ln{x_2}\ln{x_3}
                 - \frac{3}{2}\ln^2{x_3} - \frac{\pi^2}{3} -1 \right ]H^{(0)}.
\label{eq:boxp}
\eeq
The three sub-diagrams Figs.~\ref{fig:fig7}(c,d,f) all generate the double logarithms, because the two end-points of the radiated gluon
is attached to the external lines, which could  result in the soft and collinear singularities.
The Fig.~\ref{fig:fig7}(b) contains only the collinear divergence in the $l \parallel p_2$ region because one end-point of the radiated
gluon is attached to the internal gluon.

For the remaining IR singularities generated in Figs.~(\ref{fig:fig5},\ref{fig:fig6},\ref{fig:fig7}),
we can sort them into two groups
as shown in Eqs.~(\ref{eq:IR1},\ref{eq:IR2}) by using the phase space splicing
method \cite{prd65-094032}:
one is from the region $l \parallel p_2$ and the other is from the region $l \parallel p_3$.
\beq
G^{(1)}_{\rm IR1} &=& \frac{\alpha_s C_F}{4\pi} \left[ -2\ln{x_2}\ln{\delta_2} - 4\ln{\delta_2} \right] H^{(0)},
\label{eq:IR1} \\
G^{(1)}_{\rm IR2} &=& \frac{\alpha_s C_F}{8\pi} \left[ -2\ln{x_3}\ln{\delta_3} - 4\ln{\delta_3} \right] H^{(0)}.
\label{eq:IR2}
\eeq

As for the UV divergences, they are forbidden for the Feynman diagrams in Fig.~\ref{fig:fig7}
from the surface divergence analysis.
The UV divergences in the NLO quark level diagrams in Figs.~(\ref{fig:fig5},\ref{fig:fig6})
can be summed up and written in the form of
\beq
\frac{\alpha_{s}}{4 \pi}\left (11 - \frac{2}{3} N_{f} \right ) \frac{1}{\epsilon}.
\label{eq:quarkUV}
\eeq
Such UV divergence is the same one as that appeared in the  pion electromagnetic
form factors \cite{prd83-054029,prd89-054015}.

\subsection{Convolutions of the NLO Wave Functions With the LO Hard Kernel}

As argued in Refs.~\cite{prd64-014019,prd67-034001,epjc40-395,prd89-054015}, the IR divergences of the NLO corrections from the
quark level Feynman diagrams in Figs.~(\ref{fig:fig5},\ref{fig:fig6},\ref{fig:fig7}) can be absorbed into the non-perturbative
wave functions which are universal.
Based on this argument, we will make a convolution of the NLO wave functions with  the LO hard kernel $H^{(0)}$,
and find that the resultant IR part should cancel the IR divergences appeared in the NLO amplitude
$G^{(1)}_{\rm IR1}$ and $G^{(1)}_{\rm IR2}$ as given in Eqs.~(\ref{eq:IR1},\ref{eq:IR2}).
The twist-2 part of the initial pion wave function $\Phi_{\pi,A}(x_2,k_{2T};x'_2,k'_{2T})$ and the twist-3 part of the
final state pion wave function $\Phi_{\pi,P}(x'_3,k'_{3T};x_3,k_{3T})$ can be defined by the non-local matrix
elements \cite{prd64-014019,prd67-034001,epjc40-395,prd89-054015},
\beq
\Phi_{\pi,A}(x_2,k_{2T};x'_2,k'_{2T}) &=&
\int \frac{d y^-}{2 \pi} \frac{d^2 y_T}{(2 \pi)^2} e^{-i x'_2 P^+_2 y^- + i \textbf{k}'_{2T} \cdot \textbf{y}_T} \non
&& \hspace{-2cm}\cdot <0\mid\overline{q}(y) \gamma_5 \nsl_{-} W_y(n_1)^{\dag}I_{n_1;y,0} W_0(n_1) q(0) \mid \overline{u}(P_2 - k_2) d(k_2)>, \label{eq:nlowfpion1A}\\
\Phi_{\pi,P}(x'_3,k'_{3T};x_3,k_{3T}) &=& \int \frac{d z^+}{2 \pi}\frac{d^2 z_T}{(2 \pi)^2} e^{-i x'_3 P^-_3 z^+
+ i \textbf{k}'_{3T} \cdot \textbf{z}_T} \non
&&\hspace{-2cm}\cdot <0\mid\overline{q}(z) W_z(n_2)^{\dag} I_{n_2;z,0}
W_0(n_2) \gamma_5 q(0)  \mid u(P_3 - k_3) \overline{d}(k_3)>, \label{eq:nlowfpion2P}
\eeq
where $y = (0, y^-, \textbf{y}_T)$ and $z = (z_+, 0, \textbf{z}_T)$ are the light-cone coordinates of the anti-quark field
$\bar{q}$, $W_y(n_1)$ and $W_y(n_2)$ with the choice of $n^2_i \neq 0$ to avoid the light-cone
singularity\cite{prd85-074004,jhep0601-067,liLC2014}
are the Wilson line integrals:
\beq
W_y(n_1) &=& {\cal P}\; \exp [-i g_s \int^{\infty}_{0} d \lambda n_1 \cdot A(y + \lambda n_1)],
\label{eq:wl01} \\
W_z(n_2) &=& {\cal  P}\; \exp [-i g_s \int^{\infty}_{0} d \lambda n_2 \cdot A(z + \lambda n_2)],
\label{eq:wl02}
\eeq
where the symbol ${\cal P}$ denotes the path ordering operator.

\begin{figure}[tb]
\centering
\vspace{-2cm}
\leftline{\epsfxsize=12cm\epsffile{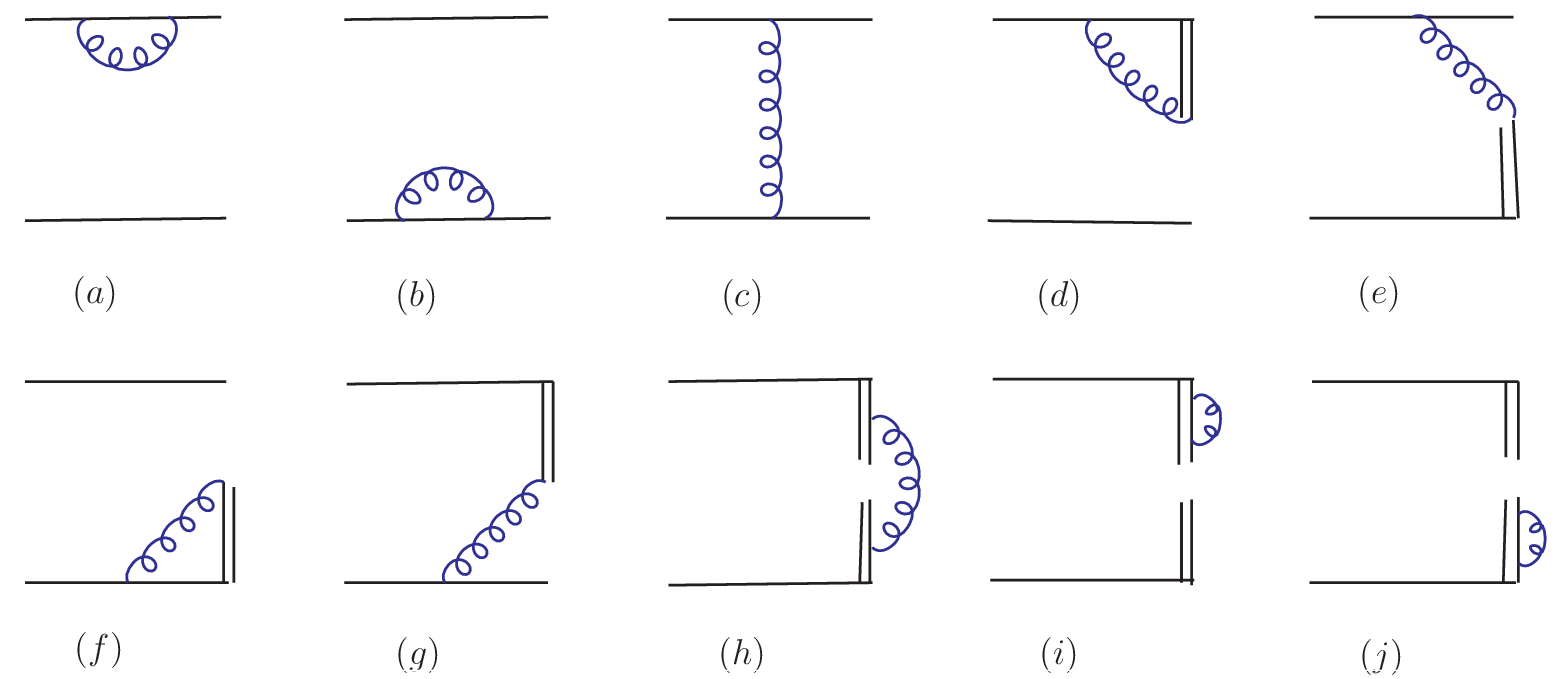}}
\vspace{-10cm}
\caption{The effective $O(\alpha_s)$ diagrams for the twist-2 initial $\pi$ meson wave functions.}
\label{fig:fig8}
\end{figure}
We firstly consider the convolutions of the $\calo(\alpha_s)$ twist-2 initial pion wave functions
$\Phi^{(1)}_{\pi,A,i}$, as shown in Fig.~\ref{fig:fig8}, with the $\calo(\alpha_s)$ hard kernel
$H^{(0)}$ in Eq.~(\ref{eq:lothka1}),
\beq
\Phi^{(1)}_{\pi,A} \otimes H^{(0)} \equiv \sum^h_{i=a} \int dx'_2 d^2 \textbf{k}'_{2T}
\Phi^{(1)}_{\pi,A,i}(x_2,\textbf{k}_{2T};x'_2,\textbf{k}'_{2T}) H^{(0)}(x'_2,\textbf{k}'_{2T};x_3,\textbf{k}_{3T}).
\label{eq:conv1}
\eeq

The reducible effective diagram Fig.~\ref{fig:fig8}(c) carry all the NLO contributions from the reducible
diagrams Fig.~\ref{fig:fig7}(a), so we can also set it's contribution to be zero safely.
The convolutions of the NLO initial wave functions $\Phi^{(1)}_{\pi,A,i}$ and the LO hard kernel $H^{(0)}$ are
summarized as
\beq
&&\Phi^{(1)}_{\pi,A,a} \otimes H^{(0)} = -\frac{\alpha_s C_F}{8 \pi} \left [\frac{1}{\epsilon} +
                   \ln{\frac{4 \pi \mu^2_f}{\delta_{2} Q^2 e^{\gamma_E}}} + 2 \right ] H^{(0)}, \non
&&\Phi^{(1)}_{\pi,A,b} \otimes H^{(0)} = -\frac{\alpha_s C_F}{8 \pi} \left [\frac{1}{\epsilon} +
                   \ln{\frac{4 \pi \mu^2_f}{\delta_{2} Q^2 e^{\gamma_E}}} + 2\right ] H^{(0)}, \non
&&\Phi^{(1)}_{\pi,A,c} \otimes H^{(0)} \equiv 0, \non
&&\Phi^{(1)}_{\pi,A,d} \otimes H^{(0)} = \frac{\alpha_s C_F}{4 \pi} \left [\frac{1}{\epsilon} +
                   \ln{\frac{4 \pi \mu^2_f}{\xi^{2}_{2} e^{\gamma_E}}} - \ln^{2}{(\delta_{2}r_{Q2})}
                   - 2 \ln{(\delta_{2} r_{Q2})} - \frac{ \pi^{2}}{3} +2 \right ] H^{(0)}, \non
&&\Phi^{(1)}_{\pi,A,e} \otimes H^{(0)} = \frac{\alpha_s C_F}{4 \pi} \left [\ln^2{(\frac{\delta_{2} r_{Q2}}{x_{2}})}
                   + \pi^{2} \right ] H^{(0)}, \non
&&\Phi^{(1)}_{\pi,A,f} \otimes H^{(0)} = \frac{\alpha_s C_F}{4 \pi} \left [\frac{1}{\epsilon} +
                   \ln{\frac{4 \pi \mu^2_f}{\xi^{2}_{2} e^{\gamma_E}}}
                   - \ln^{2}{(\frac{\delta_{2} r_{Q2}}{x^{2}_{2}})}
                   - 2 \ln{(\frac{\delta_{2} r_{Q2}}{x^{2}_{2}})} - \frac{ \pi^{2}}{3} +2 \right ] H^{(0)}, \non
&&\Phi^{(1)}_{\pi,A,g} \otimes H^{(0)} = \frac{\alpha_s C_F}{4 \pi} \left [\ln^{2}{(\frac{\delta_{2} r_{Q2}}{x^{2}_{2}})}
                   - \frac{\pi^{2}}{3} \right ] H^{(0)}, \non
&&(\Phi^{(1)}_{\pi,A,h} + \Phi^{(1)}_{\pi,A,i} + \Phi^{(1)}_{\pi,A,j}) \otimes H^{(0)} = \frac{\alpha_s C_F}{2 \pi}
                   \left [\frac{1}{\epsilon} + \ln{\frac{4 \pi \mu^2_f}{Q^2 e^{\gamma_E}}} - \ln{\delta_{23}} \right ] H^{(0)},
\label{eq:nloedi}
\eeq
where $r_{Q2} = Q^2 / \xi^2_2$ and the scale $\xi^2_2 \equiv 4 (n_1 \cdot p_2)^2/ |n^2_1| = Q^2 |n^-_1/n^+_1|$
are introduced to regularize the light corn singularity.
We can find that the double logarithms only generated from the effective diagrams without the loop momentum $l$
flowing into the LO hard kernel, such as the case in Figs.~\ref{fig:fig8}(d) and 8(f), because the effective
diagrams with the soft loop momentum flowing into the LO hard kernel are highly suppressed by the dynamics.
These double logarithms are canceled each other completely, resulting in single logarithms only.
These single logarithms will be canceled by the IR singularity as given in Eq.~(\ref{eq:IR1}) from the NLO quark
level diagrams.

\begin{figure}[tb]
\centering
\vspace{-2cm}
\leftline{\epsfxsize=12cm\epsffile{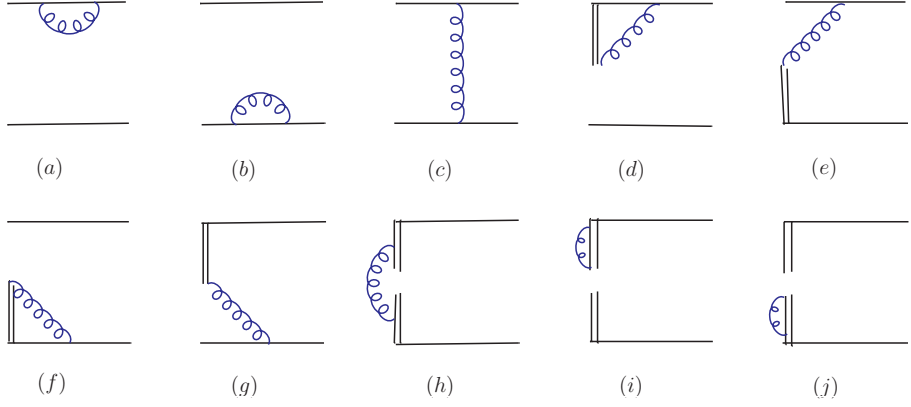}}
\vspace{-10cm}
\caption{The effective $O(\alpha_s)$ diagrams for the twist-3 final $\pi$ meson wave functions.}
\label{fig:fig9}
\end{figure}
The remaining convolutions to be treated are those between the $\calo(\alpha_s)$ hard kernel $H^{(0)}$ and the
$\calo(\alpha_s)$ twist-3 final state pion wave functions $\Phi^{(1)}_{\pi,P,i}$ as shown in Fig.~\ref{fig:fig9}.
\beq
H^{(0)} \otimes \Phi^{(1)}_{\pi,P} \equiv \sum^h_{i=a} \int dx'_3 d^23 \textbf{k}'_{3T}
H^{(0)}(x_2,\textbf{k}_{2T};x'_3,\textbf{k}'_{3T}) \Phi^{(1)}_{\pi,P,i}(x'_3,\textbf{k}'_{3T};x_3,\textbf{k}_{3T}).
\label{eq:conv2}
\eeq
We can also set the convolution of the $H^{(0)}$ and Fig.~\ref{fig:fig9}(c) zero with the same reason as for
the Fig.~\ref{fig:fig8}(c). Then all the convolutions of the effective diagrams in Fig.~\ref{fig:fig9} read as
\beq
&&H^{(0)} \otimes \Phi^{(1)}_{\pi,P,a} =-\frac{\alpha_s C_F}{8 \pi} \left [\frac{1}{\epsilon} +
                  \ln{\frac{4 \pi \mu^2_f}{\delta_{3} Q^2 e^{\gamma_E}}} + 2 \right ] H^{(0)}, \non
&&H^{(0)} \otimes \Phi^{(1)}_{\pi,P,b} = -\frac{\alpha_s C_F}{8 \pi}  \left[\frac{1}{\epsilon} +
                  \ln{\frac{4 \pi \mu^2_f}{\delta_{3} Q^2 e^{\gamma_E}}} + 2\right] H^{(0)}, \non
&&H^{(0)} \otimes \Phi^{(1)}_{\pi,P,c} \equiv 0, \non
&&H^{(0)} \otimes \Phi^{(1)}_{\pi,P,d} =  \frac{\alpha_s C_F}{8 \pi}  \left[\frac{1}{\epsilon} +
                  \ln{\frac{4 \pi \mu^2_f}{\xi^{2}_{3} e^{\gamma_E}}} - \ln^{2}{(\delta_{3}r_{Q3})}
                  - 2 \ln{(\delta_{3}r_{Q3})} - \frac{ \pi^{2}}{3} +2\right] H^{(0)},\non
&&H^{(0)} \otimes \Phi^{(1)}_{\pi,P,e} =  \frac{\alpha_s C_F}{8 \pi}  \left[\ln^2{(\frac{\delta_{3}r_{Q3}}{x_{3}})}
                  + \pi^{2}\right] H^{(0)}, \non
&&H^{(0)} \otimes \Phi^{(1)}_{\pi,P,f} =  \frac{\alpha_s C_F}{8 \pi}  \left[\frac{1}{\epsilon} +
                  \ln{\frac{4 \pi \mu^2_f}{\xi^{2}_{3} e^{\gamma_E}}} - \ln^{2}{(\frac{\delta_{3}r_{Q3}}{x^{2}_{3}})}
                  - 2 \ln{(\frac{\delta_{3}r_{Q3}}{x^{2}_{3}})} - \frac{ \pi^{2}}{3} +2\right] H^{(0)}, \non
&&H^{(0)} \otimes \Phi^{(1)}_{\pi,P,g} =  \frac{\alpha_s C_F}{8 \pi}  \left[\ln^{2}{(\frac{\delta_{3}r_{Q3}}{x^{2}_{3}})}
                  - \frac{\pi^{2}}{3}\right] H^{(0)}, \non
&&H^{(0)} \otimes (\Phi^{(1)}_{\pi,P,h} + \Phi^{(1)}_{\pi,P,i} + \Phi^{(1)}_{\pi,P,j}) =  \frac{\alpha_s C_F}{4 \pi}
                  \left [\frac{1}{\epsilon} + \ln{\frac{4 \pi \mu^2_f}{Q^2 e^{\gamma_E}}} - \ln{\delta_{23}}\right] H^{(0)},
\label{eq:nloedf}
\eeq
where $r_{Q3} = Q^2 / \xi^2_3$ with the scale $\xi^2_3 \equiv 4(n_2 \cdot p_3)^2/ |n^2_2| = Q^2 |n^+_2/n^-_2|$.
The double logarithms in Eq.~(\ref{eq:nloedf}) are also canceled each other as the case in Eq.~(\ref{eq:nloedi}),
and the remaining single logarithms can also been canceled by the IR singularity in Eq.~(\ref{eq:IR2}).
When compared with the convolutions of the irreducible diagrams in Figs.~\ref{fig:fig8}(d,e,f,g), there is an
additional factor $1/2$ for those of the irreducible diagrams Fig.~\ref{fig:fig9}(d,e,f,g),
since the twist-3 final state wave functions $\Phi^{(1)}_{\pi,P,i}$ have different spin structure from
the twist-2 initial state wave function $\Phi^{(1)}_{\pi,A,i}$.

\subsection{The NLO Hard Kernel}

The $\kt$ factorization theorem states that the NLO hard kernel can be
obtained by taking the difference of the
NLO quark level diagrams and the convolutions of LO hard kernel with NLO wave
functions \cite{prd64-014019,prd67-034001,epjc40-395,prd89-054015}, i.e.,
\beq
&&H^{(1)}(x_2,\textbf{k}_{2T};x_3,\textbf{k}_{3T}) = G^{(1)}(x_2,\textbf{k}_{2T};x_3,\textbf{k}_{3T}) \non
&&~~~~~~~~~~~~~~- \sum^{h}_{i=a} \int dx'_2 d^2 \textbf{k}'_{2T} \Phi^{(1)}_{\pi,A,i}(x_2,\textbf{k}_{2T};x'_2,\textbf{k}'_{2T})
H^{(0)}(x'_2,\textbf{k}'_{2T};x_3,\textbf{k}_{3T}) \non
&&~~~~~~~~~~~~~~- \sum^{h}_{i=a} \int dx'_3 d^2 \textbf{k}'_{3T} H^{(0)}(x_2,\textbf{k}_{2T};x'_3,\textbf{k}'_{3T})
\Phi^{(1)}_{\pi,P,i}(x'_3,\textbf{k}'_{3T};x_3,\textbf{k}_{3T}).
\label{eq:nlohk}
\eeq

Besides the contributions from the reducible diagrams, we here sum up all  $G^{(1)}_i$ as given in
Eqs.~(\ref{eq:self},\ref{eq:vertex},\ref{eq:boxp}) to obtain the NLO corrections $G^{(1)}$
from the quark level diagrams in Figs.~(\ref{fig:fig5},\ref{fig:fig6},\ref{fig:fig7}) for $N_f = 6$ and find the result,
\beq
G^{(1)} &=& \frac{\alpha_s C_F}{8 \pi} \Biggl [ \frac{29}{2}
            \left ( \frac{1}{\epsilon} + \ln{\frac{4 \pi \mu^2}{Q^2 e^{\gamma_E}}} \right )
            - 4 \ln{\delta_{2}(\ln{x_{2}} + 1) - 2 \ln{\delta_{3}}}(\ln{x_{3}} + 1) - \frac{1}{4} \ln^2{\delta_{23}} \non
        &&  + \frac{9}{4} \ln{\delta_{23}} + \frac{1}{2} \ln{x_2}\ln{x_3} + \frac{5}{8} \ln^2{x_3}
            - \frac{43}{4}\ln{x_2} + \frac{9}{4}\ln{x_3} - \frac{267 \pi^{2}}{48} + \frac{65}{2} \Biggr ] H^{(0)}.\non
\label{eq:nloqd}
\eeq
By summing up all convolutions as listed in Eqs.~(\ref{eq:nloedi},\ref{eq:nloedf}) for Figs.~(\ref{fig:fig8},\ref{fig:fig9})
without the reducible diagrams, we find the total result:
\beq
\Phi^{(1)}_{\pi,A} \otimes H^{(0)} &=& \frac{\alpha_s C_F}{4 \pi}
     \Bigl [ \frac{4}{\epsilon} + 4 \ln{\frac{4 \pi}{e^{\gamma_E}}} + 4 \ln{\frac{\mu^2_f}{Q^{2}}}
             - 2\ln{(\delta_2 r_{Q2})}(\ln{x_{2}} + 2) \non
     &&      + \ln^2{x_2} - 2\ln{\delta_{23}} + 4\ln{(x_2 r_{Q2})} - \pi^{2} + 4 \Bigr ] H^{(0)},
\label{eq:nloedI} \\
H^{(0)} \otimes \Phi^{(1)}_{\pi,P} &=& \frac{\alpha_s C_F}{8 \pi}
     \Bigl [ \frac{4}{\epsilon} + 4 \ln{\frac{4 \pi}{e^{\gamma_E}}} + 4 \ln{\frac{\mu^2_f}{Q^{2}}}
             - 2 \ln{(\delta_3 r_{Q3})}(\ln{x_{3}} + 2) \non
     &&      + \ln^2{x_3} - 2\ln{\delta_{23}} + 4\ln{(x_3 r_{Q3})} - \pi^{2} + 4 \Bigr ] H^{(0)}.
\label{eq:nloedF}
\eeq
The UV divergence in Eq.~(\ref{eq:nloqd}), which would determine the renormalization-group (RG)
evolution of the strong coupling constant $\alpha_s$, is the same one as that in the pion electromagnetic form
factor as given in Refs.~\cite{prd83-054029,prd89-054015}.
The bare coupling constant $\alpha_s$ in Eqs.~(\ref{eq:nloqd},\ref{eq:nloedI},\ref{eq:nloedF}) can be rewritten as
\beq
\alpha_s = \alpha_s(\mu_f) + \delta Z(\mu_f) \alpha_s(\mu_f), \label{eq:renormcc}
\eeq
with the counter-term $\delta Z(\mu_f)$ defined in the modified minimal subtraction scheme($\overline{MS}$).
We can insert the $\alpha_s$ in Eq.~(\ref{eq:renormcc}) into
Eqs.~(\ref{eq:lothka1},\ref{eq:nloqd},\ref{eq:nloedI},\ref{eq:nloedF})
to regularize the UV poles in Eq.~(\ref{eq:nlohk}) through the term $\delta Z (\mu_f) H^{(0)}$,
and then the UV poles in Eqs.~(\ref{eq:nloedI},\ref{eq:nloedF}) are regulated by the
counter-term of the quark field and by an additional counter-term in Eq.~(\ref{eq:renormcc}).

One should be careful that the internal quark with the tiny momentum fraction $x_2$ would be on-shell,
which would then generate an additional double logarithm $\ln^2{x_2}$, so we must subtract this jet function
as described in Eq.~(\ref{eq:jetfunction}) to obtain the real NLO hard kernel.
\beq
J^{(1)} H^{(0)} &=& - \frac{1}{2} \frac{\alpha_s(\mu_f) C_F}{4 \pi}
\left [\ln^2{x_2} + \ln{x_2} + \frac{\pi^2}{3} \right ] H^{(0)}.
\label{eq:jetfunction}
\eeq

After renormalizing the UV divergences and subtracting the jet function, one can obtain the NLO hard kernel for Fig.~\ref{fig:fig1}(a)
by combing the results as given previously in Eqs.~(\ref{eq:nlohk},\ref{eq:nloqd},\ref{eq:nloedI},\ref{eq:nloedF}) together:
\beq
H^{(1)}(x_i,\mu,\mu_f,Q^2) &\equiv& F^{(1)}(x_i,\mu,\mu_f,Q^2) H^{(0)},
\label{eq:nloHK}
\eeq
with
\beq
F^{(1)}(x_i,\mu,\mu_f,Q^2)&=& \frac{\alpha_s(\mu_f) C_F}{8 \pi}
        \Biggl [ \frac{21}{2} \ln{\frac{\mu^2}{Q^2}} - 8 \ln{\frac{\mu^2_f}{Q^2}}
              - \frac{1}{4}\ln^{2}{\delta_{23}} + \frac{33}{4} \ln{\delta_{23}} + \frac{1}{2} \ln{x_2}\ln{x_3}  \non
    &&  - \frac{3}{8} \ln^2{x_3} - \ln^2{x_2} - \frac{71}{4} \ln{x_2}
              - \frac{7}{4} \ln{x_3} - \frac{107}{48} \pi^{2} + \frac{41}{2} \Biggr ],
\label{eq:nlof1-1a}
\eeq
here one has made the choice for $r_{Q2}=r_{Q3} \equiv 1$ as in Refs.~\cite{prd83-054029,prd89-054015}.

\subsection{Numerical results and discussions}

In this subsection we will calculate the NLO corrections to the space-like
scalar pion form factor in the $k_T$ factorization theorem numerically.
From the expression of the NLO hard kernel $H^{(1)}(x_i,\mu,\mu_f,Q^2)$ as
given in Eq.~(\ref{eq:nloHK}),
one can define the space-like scalar pion form factor for Fig.~\ref{fig:fig1}(a)
up to NLO as the form of
\beq
Q^2F(Q^2)|&&_{\rm NLO} = 8 \pi m_{0\pi} C_F Q^4 \int{dx_2 dx_3} \int{b_2 db_2 b_3 db_3}
              ~ \alpha_s(t) \cdot e^{-2 S_{\pi}(t)} \cdot h(x_2,x_3,b_2,b_3) \non
      &&  \hspace{-1cm}\cdot \Bigl \{2 \phi^{A}_{\pi}(x_{2}) \phi^{P}_{\pi}(x_{3}) S_t(x_3)
      \left[ 1+ F^{(1)}(x_i,\mu,\mu_f,Q^2) \right]
+ x_2 \left[ \phi^{P}_{\pi}(x_{2}) - \phi^{T}_{\pi}(x_{3}) \right]
\phi^{A}_{\pi}(x_{3})\Bigr\}, \non\label{eq:nloff} \eeq where the
function $F^{(1)}(x_i,\mu,\mu_f,Q^2)$ describes the NLO contribution
to the space-like scalar pion form factor and has been defined in
Eq.~(\ref{eq:nlof1-1a}).

Since the initial and final state meson are the same pion meson,
which is a $q\bar{q}$ bound state and also a Nambu-Goldstone boson,
then there is an exchange symmetry of the momentum fractions for
the two sub-diagrams in Fig.~\ref{fig:fig1},
as we have demonstrated in Section.~II.
This symmetry imply that the NLO correction $F^{(1)}_{b}(x_i,\mu,\mu_f,Q^2)$ to the dominant first term
proportional to $\phi^{A}_{\pi}(x_{3}) \phi^{P}_{\pi}(x_{2})$ in $H^{(0)}_{b}(x_{2},x_{3},Q^2)$
in Eq.~(\ref{eq:lothkb}) can be obtained from $F^{(1)}(x_i,\mu,\mu_f,Q^2)$ in Eq.~(\ref{eq:nlof1-1a})
by simple replacements of $x_2 \leftrightarrow x_3$, and is of the form
\beq
F^{(1)}_{b}(x_i,\mu,\mu_f,Q^2) &=& \frac{\alpha_s(\mu_f) C_F}{8 \pi}
        \Biggl [ \frac{21}{2} \ln{\frac{\mu^2}{Q^2}} - 8 \ln{\frac{\mu^2_f}{Q^2}}
              - \frac{1}{4}\ln^{2}{\delta_{23}} + \frac{33}{4} \ln{\delta_{23}} \non
    && \hspace{-3cm} + \frac{1}{2} \ln{x_3}\ln{x_2} - \frac{3}{8} \ln^2{x_2} - \ln^2{x_3} - \frac{71}{4} \ln{x_3}
              - \frac{7}{4} \ln{x_2} - \frac{107}{48} \pi^{2} + \frac{41}{2} \Biggr ].
\label{eq:nlof1-1b}
\eeq
while the space-like scalar pion form factor from Fig.~\ref{fig:fig1}(b) up
to NLO level can be written as the form of
\beq
Q^2F(Q^2)|_{\rm b, NLO} &=&8 \pi m_{0\pi} C_F Q^4 \int{dx_2 dx_3} \int{b_2 db_2 b_3 db_3}
              ~ \alpha_s(t) \cdot e^{-2 S_{\pi}(t)} \cdot h(x_2,x_3,b_2,b_3) \non
      &&  \hspace{-2cm}\cdot \Bigl \{2 \phi^{A}_{\pi}(x_{3}) \phi^{P}_{\pi}(x_{2}) S_t(x_2)
      \left[ 1+ F^{(1)}_b(x_i,\mu,\mu_f,Q^2) \right]
+ x_3 \left[ \phi^{P}_{\pi}(x_{3}) - \phi^{T}_{\pi}(x_{3}) \right]
\phi^{A}_{\pi}(x_{2})\Bigr\}.\non\label{eq:nloff2} \eeq Explicit
analytical calculations also confirmed this exchanging symmetry
directly.

\begin{figure}[tb]
\centering
\begin{minipage}{0.45\textwidth}
\centerline{\epsfxsize=8cm\epsffile{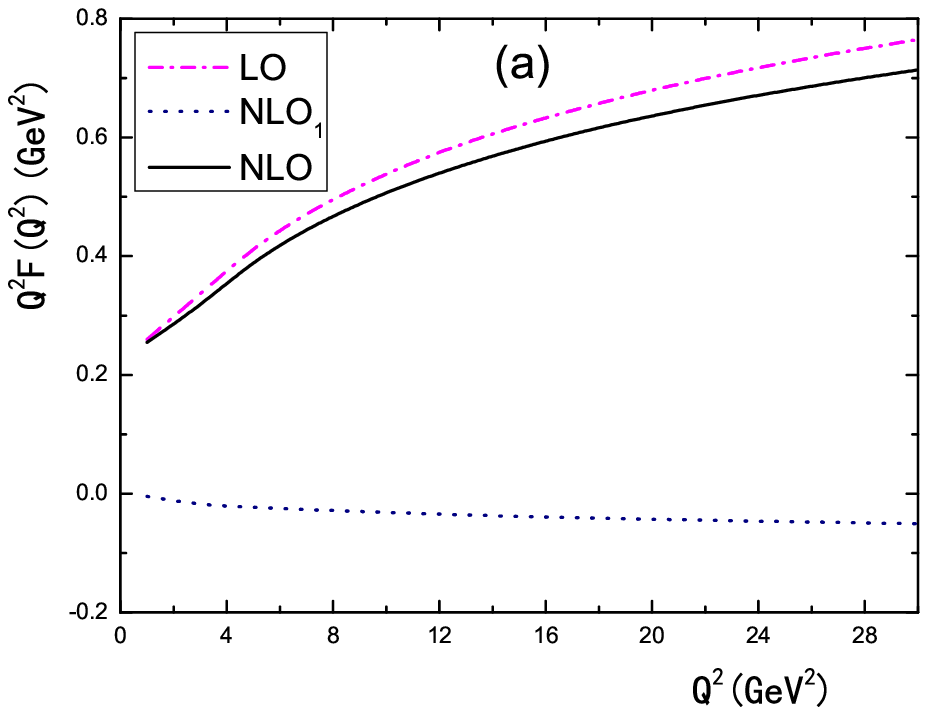}}
\end{minipage}
\begin{minipage}{0.45\textwidth}
\centerline{\epsfxsize=8cm\epsffile{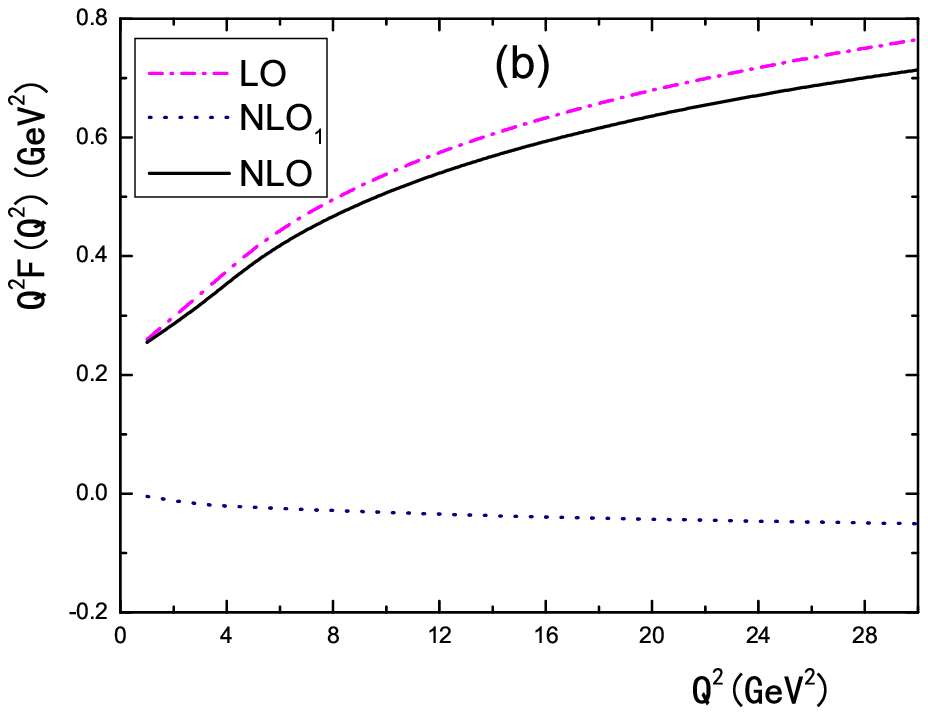}}
\end{minipage}
\vspace{-0.5cm}
\caption{The pQCD predictions for the space-like scalar pion form factor $Q^2F(Q^2)$ for
Fig.~\ref{fig:fig1}(a) and for Fig.~\ref{fig:fig1}(b). }
\label{fig:fig10}
\end{figure}

In Fig.~\ref{fig:fig10}, we plot the $Q^2$-dependence of the pQCD predictions for the form factor
$Q^2 F(Q^2)$.  The Fig.~\ref{fig:fig10}(a) and Fig.~\ref{fig:fig10}(b) shows the result from
Fig.~\ref{fig:fig1}(a) and Fig.~\ref{fig:fig1}(b), respectively.
The upper dot-dashed and lower dotted curve shows the LO contribution and the NLO correction respectively,
while the solid curve refers to the total pQCD predictions after the inclusion of the NLO corrections.
From the numerical results as illustrated in Fig.~\ref{fig:fig10}, we find the following points:
\begin{enumerate}
\item
As shown by the dots line in Fig.~\ref{fig:fig10}, the NLO correction to the LO pQCD prediction for $Q^2 F(Q^2)$
is negative in sign and very small in magnitude in the whole considered region of $Q^2$.
The inclusion of the NLO corrections can produce a small decrease, less than $8\%$ in magnitude,
to the LO result in the region of $1\leq Q^2 \leq 30$ GeV$^2$.

\item
As illustrated by the curves in Fig.~\ref{fig:fig10}(a,b), the LO and NLO contributions to the form factor  $Q^2 F(Q^2)$  from
Fig.~\ref{fig:fig1}(a) and Fig.~\ref{fig:fig1}(b) are indeed identical, which is what we expect based on
the exchanging symmetry as discussed in Section II.
\end{enumerate}

\section{NLO corrections and effects on $B \to \pi\pi$ decays}

In this section we will extend our calculations for the NLO correction to the LO space-like scalar pion form factor
to the case in the time-like range by the analytical continuation, and then revisit the puzzled
$B \to \pi\pi$ decays with the inclusion of this new NLO correction by employing the pQCD factorization approach.

\subsection{NLO corrections to the time-like scalar pion form factor}

With the NLO space-like scalar pion hard kernel in Eq.~(\ref{eq:nlof1-1a}) and the analytical continuation relation,
we can obtain the NLO hard amplitude for the time-like scalar pion form factor in the $\kt$ space by substituting
$-M^2_B - i \epsilon$ for the scale $Q^2$ of the factorizable annihilation process in the B meson decays,
and $-x_2(1-x_3)M^2_B + |\ktb-\ktc|^2 -i \epsilon$ for the internal gluon.
The single-b convoluted NLO time-like hard kernel can be expressed as
\beq
H'^{(1)}_{a,32}(x_i,\kt,t,M^2_B) &\equiv& F'^{(1)}_{a,32}(x_i,\kt,t,M^2_B) \cdot H'^{(0)}_{a,32},
\label{eq:nloHKTL} \\
F'^{(1)}_{a,32}(x_i,\kt,t,M^2_B) &=& \frac{\alpha_s(\mu_f) C_F}{8 \pi}
        \Bigl [ \frac{5}{2} \ln{\frac{\mu^2}{M^2_B}}
              - \frac{1}{4}\ln^{2}{\delta'_{23}} + \frac{33}{4} \ln{\delta'_{23}}
              - \ln^2{(1-x_3)} \non
    &&  - \frac{3}{8} \ln^2{x_2} + \frac{1}{2} \ln{x_2} \ln{(1-x_3)}
              - \frac{71}{4} \ln{(1-x_3)}  \non
    &&  - \frac{7}{4} \ln{x_2} - \frac{95}{48} \pi^{2} + \frac{41}{2}
              + i \pi \left (\frac{1}{2} \ln{\delta'_{23}} - \frac{23}{4}\right ) \Bigr ],
\label{eq:nlofactorTL}
\eeq
where $H'^{(0)}_{a,32}$ has been given in Eq.~(\ref{eq:lohka32}), $H'^{(1)}_{a,32}$ and $F'^{(1)}_{a,32}$
represent the corresponding NLO time-like scalar hard kernel and the NLO
correction factor with the following notation,
\beq
\ln{\delta'_{23}}=\ln{\frac{|\ktb-\ktc|^2 - x_2(1-x_3)M^2_B}{M^2_B}} + i \pi \cdot \Theta(|\ktb-\ktc|^2 - x_2(1-x_3)M^2_B).
\label{eq:deltap23}
\eeq
We can then obtain the NLO single-b convolution time-like scalar pion form factor by the Fourier
transformation of Eq.~(\ref{eq:nloHKTL}) from the $\ktc$ space to the $b_3$ space as well as the integration
over the kinematic variables. And the time-like scalar pion form factor up to NLO level
can then be written as,
\beq
F'^{(1)}_{\rm a,I} &=& - \int^1_0 dx_2 dx_3 \int^{\infty}_0 db_3 ~ \frac{C_F M^2_B m^{\pi}_0}{ \pi (1-x_3)}
       \cdot\Bigl \{  2 \phi_\pi^P(x_2) \phi_\pi^A(x_3) S_t(x_2)\; K''^{(1)}_0(z) \non
  &&\hspace{-1cm} + \Bigl [ (1 - x_3) \phi^A_{\pi}(x_2)\left (\phi^P_{\pi}(x_3) + \phi^T_{\pi}(x_3) \right )
  +2 \phi^P_{\pi}(x_2)\phi^A_{\pi}(x_3)\left (1 + F'^{(1)}_{32} \right )
                           \cdot S_{t}(x_2) \Bigr ] \cdot K_0(z)\Bigr \}\non
&& \cdot \alpha_s(\mu) \cdot \exp\left [-S_{I}(x_2,b_2;1-x_3,b_3;M_B;\mu) \right],
\label{eq:nloffI}
\eeq
where $z=i \sqrt{(1-x_3)x_2} M_B b_3$, the definition of the function $K''^{(1)}_0(z)$ is of the form
\beq
K''^{(1)}_0(z) = \left[\frac{\partial^2}{\partial{\alpha^2}} K^{(1)}_{\alpha}(z) \right]_{\alpha=0},
\label{eq:Bessel2}
\eeq
which comes from the Fourier transformation of $\ln^{2}{(|\ktb-\ktc|^2 - x_2(1-x_3)M^2_B-i \epsilon)}$,
and $\alpha$ denotes the order parameter of the modified Hankel function, and it's magnitude behaves
as $|K''^{(1)}_0(z)| \sim (1/3) \ln^2{(z)} |K_0(z)|$ when the argument $|z| \rightarrow 0$.
The NLO correction factor  $F'^{(1)}_{32}$ in Eq.~(\ref{eq:nloffI}) is of the form
\beq
F'^{(1)}_{32}(x_i,b_3,t,M^2_B) &=&  \frac{\alpha_s(\mu_f) C_F}{8 \pi}
        \left\{\frac{5}{2} \ln{\frac{\mu^2}{M^2_B}} - \frac{1}{16}\ln^{2}{\left(\frac{4x_2(1-x_3)}{M^2_B b^2_3}\right)}  \right. \non
    &&\hspace{-2cm} \left. + (\frac{33}{8} - \frac{\gamma_E}{4}) \ln{\left(\frac{4x_2(1-x_3)}{M^2_B b^2_3}\right)}
              - \ln^2{(1-x_3)}- \frac{3}{8} \ln^2{x_2} + \frac{1}{2} \ln{x_2} \ln{(1-x_3)} \right. \non
    &&\hspace{-2cm} \left. - \frac{71}{4} \ln{(1-x_3)} - \frac{7}{4} \ln{x_2} - \frac{105}{48} \pi^{2} + \frac{41}{2}
              - \frac{\gamma^2_E}{4} - \frac{33 \gamma_E}{4} + i \pi \cdot \left[\frac{5}{2} \right] \right\},
\label{nlofactor}
\eeq
where $\gamma_E$ is the Euler constant.

\subsection{NLO effects on $B \to \pi\pi$ decays}

In this subsection we will firstly show the NLO contributions to the time-like scalar pion form factor
$F'^{(1)}_{\rm a,I}$ in the $k_T$ factorization theorem numerically, and then
examine the effects of such NLO contribution on the pQCD predictions for the branching ratios
of the rare $B \to \pi \pi$ decays.

\begin{figure}[tb]
\centering
\begin{minipage}{0.45\textwidth}
\centerline{\epsfxsize=8cm\epsffile{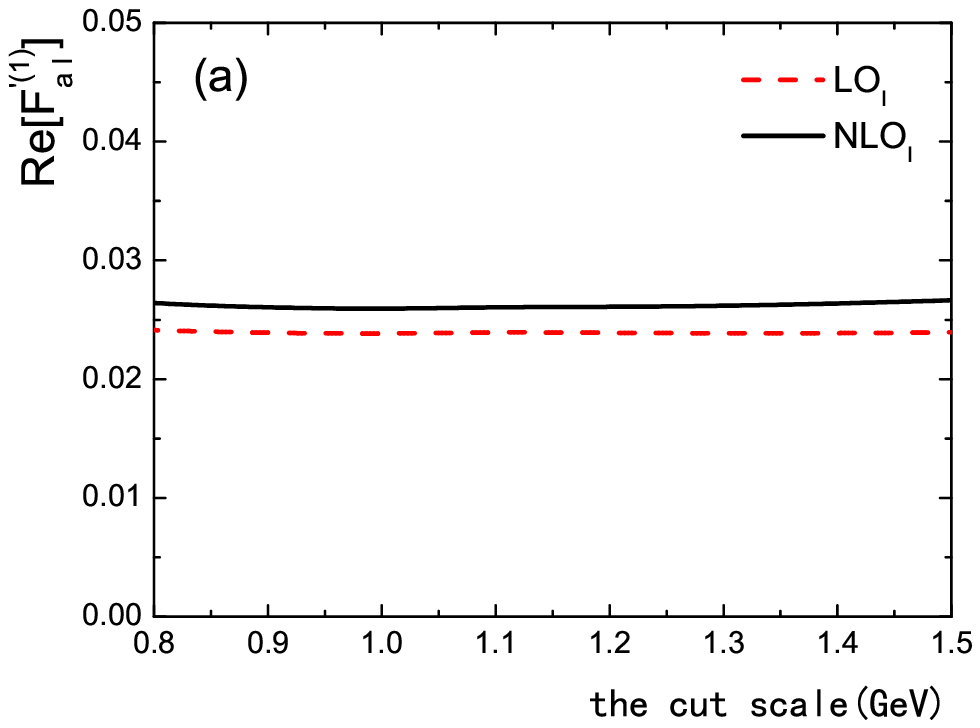}}
\end{minipage}
\begin{minipage}{0.45\textwidth}
\centerline{\epsfxsize=8cm\epsffile{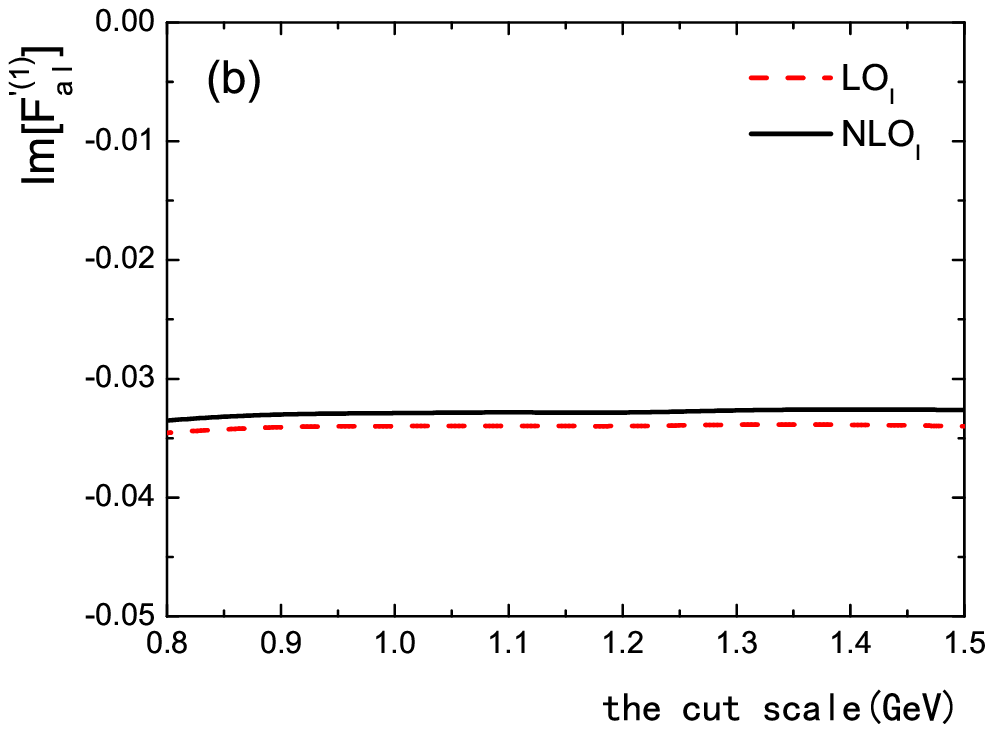}}
\end{minipage}
\vspace{0cm}
\caption{The $\mu_0$-dependence of the pQCD predictions for the
time-like scalar pion form factor $F'_{\rm a,I}$ at the LO and NLO level.
The left (right ) figure shows the real (imaginary) part of the form factor.
The short-dash and solid line shows the pQCD predictions for the form factor
$F'_{\rm a,I}$ at the LO and the NLO level, respectively.}
\label{fig:fig11}
\end{figure}

In the calculations for the LO time-like scalar pion form factor $F'^{(0)}_{\rm a}$,
we considered the cases for both the single-b and double-b convolution and found that the differences
are very small between these two different convolution methods.
Consequently, we make the calculation for the NLO form factor $F'^{(1)}_{\rm a}$ as given
in Eq.~(\ref{eq:nloffI}) by using the single-b convolution only.

In Fig.~\ref{fig:fig11}(a) and \ref{fig:fig11}(b), we show the $\mu_0$-dependence of the real and
imaginary part of the time-like scalar pion form factor $F'_{\rm a,I}$ at the LO and NLO level, respectively.
The short-dash line in Fig.~4 shows the LO contribution, while the solid line shows the form factor
after the inclusion of the NLO contribution.

\begin{figure}[tb]
\centering
\begin{minipage}{0.45\textwidth}
\centerline{\epsfxsize=8cm\epsffile{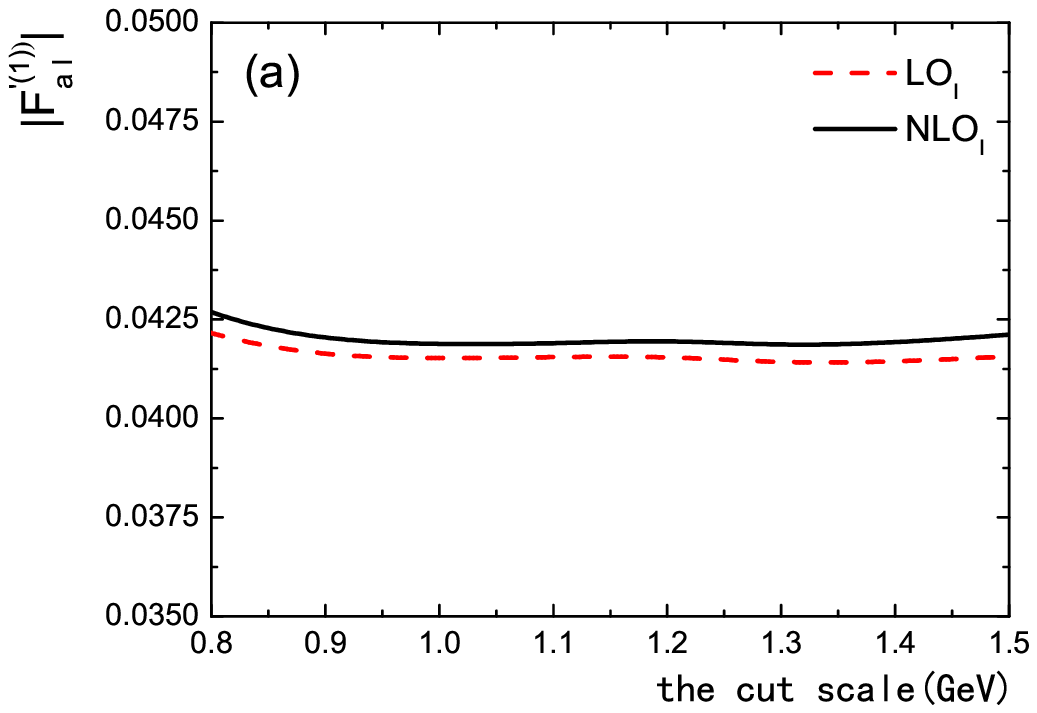}}
\end{minipage}
\begin{minipage}{0.45\textwidth}
\centerline{\epsfxsize=8cm\epsffile{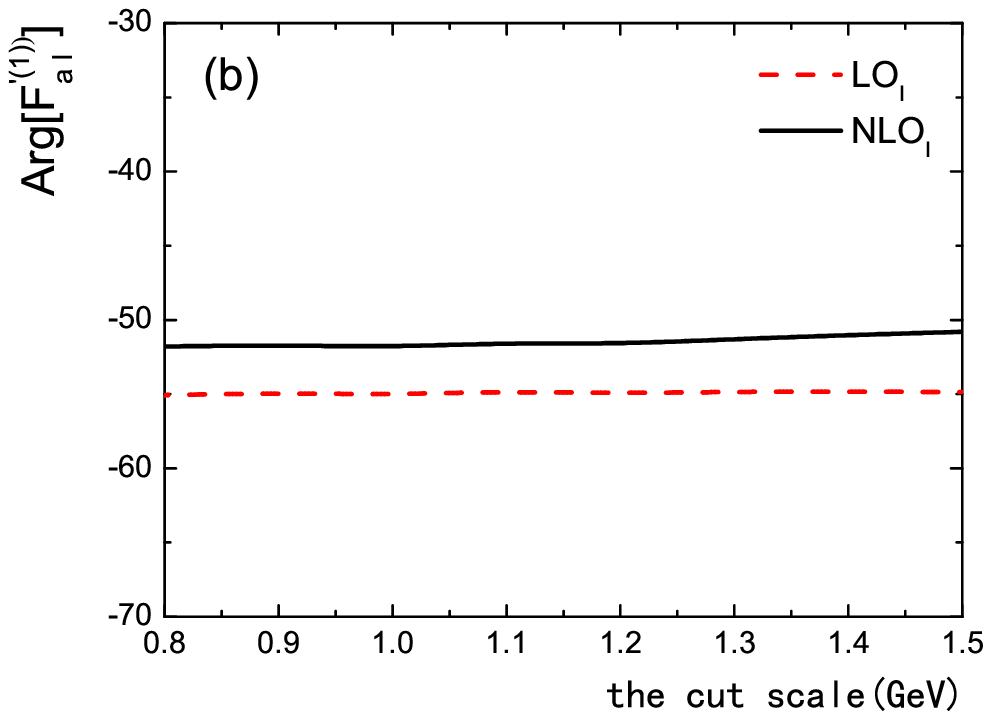}}
\end{minipage}
\vspace{0cm}
\caption{The same as in Fig.~11, but for the $\mu_0$-dependence of the
absolute value and their argument of the considered form factor $F'^{(0)}_{\rm a,I}$
and $F'^{(1)}_{\rm a,I}$.}
\label{fig:fig12}
\end{figure}

In Fig.~\ref{fig:fig12}(a) and \ref{fig:fig12}(b), however,  we show the $\mu_0$-dependence of
the pQCD predictions for the absolute values and their arguments of the time-like scalar pion
form factor $F'_{\rm a,I}$ at the LO (the short-dash line) and NLO (the solid line) level,
respectively. For fixed $\mu_0=1.0$ GeV, we have numerically
\beq
F'_{\rm a,I}&=&\left \{ \begin{array}{ll}
0.0238 - i 0.0340, & {\rm LO_{I}}, \\
0.0259 - i 0.0329, & {\rm NLO_{I}}, \\ \end{array} \right. \non
&=& \left \{ \begin{array}{ll}
0.0415\cdot\exp[-i 55.0^\circ], & {\rm LO_{I}}, \\
0.0419\cdot\exp[-i 51.8^\circ], & {\rm NLO_{I}}, \\ \end{array} \right.\label{eq:fpa-02}
\eeq

From Figs.~(\ref{fig:fig11},\ref{fig:fig12}) and the numerical results in Eq.~(\ref{eq:fpa-02}),
one can see the following points:
\begin{enumerate}
\item
The NLO part is indeed very small in size, brings little correction to both the real- and imaginary
part of the LO form factor, and is almost independent with the variation of cutoff scale $\mu_0$.

\item
The real part of the time-like scalar pion form factor is positive, while its imaginary part is negative,
which leads to a large strong phase around $-55^\circ$ and play an important role in producing large CP violation
for $B\to \pi\pi$ decays.

\end{enumerate}

\begin{table}[thb]
\begin{center}
\caption{The LO and NLO pQCD predictions for the branching rations (in unit of $10^{-6}$) of the three $B \to \pi\pi$ decays.
The last column lists the data from Refs.\cite{bf-1406,petric-201407}. For details, see text.}
\label{brpi}
\vspace{0.2cm}
\begin{tabular}{c |c| c | c| c| l}
\hline\hline
Channel~~~             &  LO   & NLO$_0$ \cite{prd90-014029}     & NLO  & QCDF\cite{npb675} & Data    \\
\hline
$Br(B^0 \to \pi^+ \pi^-)$      &$6.87$   &  $7.67$     &$7.69^{+3.27}_{-2.67}$    &$8.9$   & $5.11 \pm 0.22$  \\
$Br(B^+ \to \pi^+ \pi^0)$      &$3.54$   &  $4.27$     &$4.27^{+1.85}_{-1.47}$    &$6.0$   & $5.38^{+0.35}_{-0.34}$ \\
$Br(B^0 \to \pi^0 \pi^0)$      &$0.12$   &  $0.23$     &$0.24^{+0.09}_{-0.07}$    &$0.3$   & $0.9 \pm 0.12$ \\
\hline\hline
\end{tabular}
\end{center}
\end{table}

In the pQCD factorization approach, the NLO contribution to $B \to \pi\pi$ decays
from the factorizable annihilation diagrams are described by the time like scalar
pion form factor $F'^{(1)}_{\rm a,I}$, in other words, it is the NLO "annihilation correction".
After the inclusion of this new NLO time-like scalar pion form factor in Eq.~(\ref{eq:nloffI}),
we recalculate the three rare decays $B \to \pi \pi$ in the pQCD factorization approach
by using the pion distribution amplitudes as given in Eq.~(\ref{eq:phipi1}).
Because this newly known NLO contribution brings only a very small correction to the LO form factor as
we have elaborated in previous section, one generally expect that such new NLO contribution to
the time-like scalar pion form factor  can not change the pQCD predictions for the
$B \to \pi \pi$ decays obviously.

In the framework of the pQCD factorization approach, the LO contributions to $B\to \pi\pi$ decays
come from the emission diagrams, the hard-spectator diagrams,
the factorizable and non-factorizable annihilation diagrams as illustrated in the Fig.~1 of
Ref.~\cite{prd90-014029}.
At the NLO level, on the other hand, those currently known NLO contributions to $B\to \pi\pi$
decays include the following pieces from rather different sources:
\ben
\item
The Wilson coefficients $C_i(\mw)$ and the renormalization group evolution matrix $U(\mu,\mw,\alpha)$ at the NLO level
\cite{buras96}, as well as the strong coupling constant $\alpha_s(\mu)$
at two-loop level \cite{pdg2014}.

\item
The NLO contributions from the vertex corrections (VC), the quark-loops (QL),
and the chromo-magnetic penguin operator $O_{8g}$ (MP) as given in
Refs.~\cite{prd72-114005,npb675,o8g2003}.

\item
The NLO twist-2 and twist-3 contributions to the form factors of the $B \to \pi$ transition
as presented in Refs.~\cite{prd85-074004,prd89-094004}.

\item
The NLO contribution to the time-like scalar pion form factor $F'_{\rm a,I}$,
i.e., the NLO ``annihilation correction" to the factorizable annihilation diagrams
(see Fig.~2), evaluated firstly in this paper.

\een
The still missing NLO parts in the pQCD approach are those ${\cal O}(\alpha_s^2)$
contributions to the hard spectator diagrams and the non-factorizable annihilation diagrams.

Following the same procedure
\footnote{For the sake of simplicity, we do not show the explicit expressions of the decay amplitudes of
$B^0\to \pi^+\pi^-, \pi^0\pi^0$ and $B^+\to \pi^+\pi^0$ decays here. For relevant formulaes, one can see those
as given in Ref.~\cite{prd90-014029} explicitly. } as in Ref.~\cite{prd90-014029},
we make the numerical calculations and present the pQCD predictions for the
branching ratios of the three  $B \to \pi\pi$ decays after the inclusion of all currently known
NLO corrections in Table \ref{brpi}.
In the third column of Table \ref{brpi}, we list the NLO pQCD predictions for the branching ratios
of three decay modes as given in Ref.~\cite{prd90-014029}, where all known NLO
contributions except for the NLO contribution to the factorizable annihilation diagrams
calculated in this paper  have been taken into account.
The numerical results in the fourth column with the label "NLO", however, are obtained
with the inclusion of all currently known NLO contributions in the pQCD factorization approach.
In fifth column, we show the central values of the theoretical predictions based on
the QCDF approach \cite{npb675},
while the last column lists the data from Refs.\cite{bf-1406,petric-201407}.

From our analytical and numerical calculations for the pQCD predictions for the branching
ratios of the three $B \to \pi\pi$ decays, we have the following observations:
\begin{enumerate}
\item
The $B^+ \to \pi^+ \pi^0$ decay do not receive corrections from this new
NLO annihilation correction,  because the annihilation diagrams do not contribute to
$B^+\to \pi^+\pi^0$ decay mode.

\item
For $B^0 \to (\pi^+\pi^-,\pi^0\pi^0)$ decays, the inclusion of the NLO contribution to the factorizable
annihilation diagram can produce a very small enhancement to their branching ratios,
less than $3\%$ to the LO results.
The well-known $\pi\pi$-puzzle can not be interpreted by the inclusion of this very small
NLO contribution. This fact, on the other hand, do support the general expectation in the pQCD
factorization approach\cite{prd87-094003,prd90-014029}:
the NLO correction to the annihilation diagrams
of $B \to P P$ decays are the higher order corrections to the small quantities, and therefore
should be very small in magnitude.
\end{enumerate}

For the CP violating asymmetries of the three $B \to \pi\pi$ decays \cite{prd90-014029},
the effects due to the inclusion of the newly known NLO annihilation correction
$F'^{(1)}_{\rm a,I}$ is also very small in size and can be neglected safely.

\section{Conclusion}

In this paper, we made the first calculation for the NLO contribution to the space-like-
and time-like scalar pion form factor in the $\kt$ factorization theorem, which is in turn
the $\calo(\alpha_s^2)$ NLO correction to the factorizable annihilation diagrams for $B \to \pi\pi$
decays.
The external light quarks are all set off-shell by $k^2_{\rm iT}$ to regulate the IR divergences
which would appear in the NLO calculations.

We calculated both the NLO quark-level diagrams and the convolutions of the LO hard
kernel $H^{(0)}$ with the NLO wave functions to obtain the NLO space-like hard kernel $H^{(1)}$.
Because all quarks in this process are massless, then all the IR divergences in these
two type diagrams can be described by the logarithms $\ln^2{(k_{\rm iT})}$.
The QCD dynamics ensures that the contribution from the radiated soft gluon
is highly suppressed by $1/Q^2$ in the perturbative theory,
our LO and NLO numerical calculations confirmed this point by showing that the
double logarithms $\ln^2{(k_{\rm iT})}$ generated from
the soft kinetic region are canceled completely between the quark-level diagrams
and the effective diagrams respectively.
We then prove that all the remaining collinear divergences from the
quark-level diagrams are also canceled by those from the effective
diagrams at NLO level, which is also the basic requirement of the $\kt$
factorization theorem.

We made the numerical evaluations for the space-like scalar pion form factor
$Q^2 F(Q^2)$ up to NLO by using the full pion  DA's in the integration.
From the NLO space-like scalar pion form factor, we found the NLO time-like scalar pion
form factor  $F'^{(1)}_{\rm a,I}$ by analytical continuation, which  describes the
NLO  ${\cal O}(\alpha_s^2)$  contribution to the factorizable annihilation diagrams
for the considered $B \to \pi\pi$ decays in this paper.
By taking  the newly known NLO annihilation correction $F'^{(1)}_{\rm a,I}$ into account, we
recalculate the branching ratios of the three $B \to \pi\pi$ decays with the inclusion of all
currently known NLO contributions, and to check the effect of this NLO annihilation correction.

Based on our analytical evaluations and the numerical results, we found the following points:
\ben
\item
we completed the first analytical calculation for the NLO contribution to the space-like
and time-like scalar pion form factor in the $\kt$ factorization theorem.

\item
There is an exchanging symmetry between the LO hard kernel $H^{(0)}_{a}$ and $H^{(0)}_{b}$:
the contribution to the form factor $Q^2 F(Q^2)$ from the Fig.~\ref{fig:fig1}(a) and
Fig.~\ref{fig:fig1}(b) are indeed identical.

\item
The NLO correction to the space-like scalar pion form factor  has an opposite sign with the LO
one but is very small in magnitude, can produce at most $10\%$ decrease to
$Q^2 F(Q^2)$ in the considered $Q^2$ region.

\item
By making the analytical continuation, we found the NLO time-like scalar pion
form factor  $F'^{(1)}_{\rm a,I}$ from the space-like one, which describes the NLO
annihilation correction to the considered $B\to \pi\pi$ decays.

\item
The NLO part of the form factor $F'^{(1)}_{\rm a,I}$ is very small in size, and
is almost independent with the variation of cutoff scale $\mu_0$.
But the form factor $F'^{(1)}_{\rm a,I}$ has a large strong phase around $-55^\circ$
and can play an important role in producing large CP violation for $B\to \pi\pi$ decays.

\item
For $B^0 \to (\pi^+\pi^-,\pi^0\pi^0)$ decays, the effects of the newly known NLO contribution
to the pQCD predictions for their branching ratios are very small, less than $3\%$ in magnitude.
The well-known $\pi\pi$-puzzle can not be interpreted by the inclusion of this very small
NLO contribution.

\end{enumerate}

\section{Acknowledement}
The authors would like to thank Hsiang-nan Li and Cai-Dian L\"u for
long term collaborations and valuable discussions.
This work is supported by the National Natural Science Foundation of China under Grant No.11235005
and by the Project on Graduate Students¡¯ Education and Innovation of Jiangsu
Province under Grant No. CXZZ13-0391.


\end{document}